%% file: main.tex
\documentclass[fleqn,usenatbib]{mnras}

\usepackage{newtxtext,newtxmath}

\usepackage[T1]{fontenc}
\usepackage{ae,aecompl}

\usepackage{graphicx}	 
\usepackage{amsmath}	 

\usepackage[
    range-units=single,
    separate-uncertainty=true,
    multi-part-units=single,
    group-separator={\,}
    ]{siunitx}           

\usepackage{float}       
\usepackage{booktabs}    
\usepackage{longtable}   
\usepackage{subcaption}  
\newcommand{\numax}{\ensuremath{{\nu_\mathrm{max}}}}
\newcommand{\dnu}{\ensuremath{\Delta\nu}}
\newcommand{\metallicity}{\ensuremath{[\mathrm{M}/\mathrm{H}]}}
\newcommand{\teff}{\ensuremath{T_\mathrm{eff}}}
\newcommand{\mlt}{\ensuremath{{\alpha_\mathrm{MLT}}}}

\DeclareSIUnit\year{yr}
\DeclareSIUnit\solarmass{\ensuremath{M_\odot}}
\DeclareSIUnit\solarradius{\ensuremath{R_\odot}}
\DeclareSIUnit\solarluminosity{\ensuremath{L_\odot}}
\DeclareSIUnit\parsec{pc}
\DeclareSIUnit\dex{dex}
\DeclareSIUnit\magnitude{mag}
\DeclareSIUnit\aarcsec{as}

\defcitealias{Serenelli.Johnson.ea2017}{S17}

\title[Hierarchically modelling many stars]{%
    Hierarchically modelling \emph{Kepler} dwarfs and subgiants to improve inference of stellar properties with asteroseismology
}

\author[A. J. Lyttle et al.]{%
Alexander J. Lyttle,$^{1,2}$\thanks{\mbox{E-mail: \href{ajl573@student.bham.ac.uk}{ajl573@student.bham.ac.uk} (AJL); \href{g.r.davies@bham.ac.uk}{g.r.davies@bham.ac.uk} (GRD)}}
Guy R. Davies,$^{1,2}$\footnotemark[1]
Tanda Li,$^{1,2}$
Lindsey M. Carboneau,$^{1,2}$
\newauthor
Ho-Hin Leung,$^{3,1,2}$
Harry Westwood,$^{1,2}$
William J. Chaplin,$^{1,2}$
Oliver J. Hall,$^{4,1,2}$
\newauthor
Daniel Huber,$^{5}$
Martin B. Nielsen,$^{1,2,6}$
Sarbani Basu,$^{7}$
and Rafael A. Garc\'ia$^{8}$
\\
$^{1}$School of Physics and Astronomy, University of Birmingham, Birmingham, B15 2TT, UK\\
$^{2}$Stellar Astrophysics Centre (SAC), Department of Physics and Astronomy, Aarhus University, Ny Munkegade 120, DK-8000 Aarhus C, Denmark\\
$^{3}$School of Physics and Astronomy, University of St Andrews, North Haugh, St Andrews, KY16 9SS, UK\\
$^{4}$European Space Agency (ESA), European Space Research and Technology Centre (ESTEC), Keplerlaan 1, 2201 AZ Noordwijk, the Netherlands\\
$^{5}$Institute for Astronomy, University of Hawai'i, 2680 Woodlawn Drive, Honolulu, HI 96822, USA\\
$^{6}$Center for Space Science, NYUAD Institute, New York University Abu Dhabi, PO Box 129188, Abu Dhabi, United Arab Emirates\\
$^{7}$Department of Astronomy, Yale University, 52 Hillhouse Avenue, New Haven, CT 06511, USA\\
$^{8}$AIM, CEA, CNRS, Universit\'e Paris-Saclay, Universit\'e Paris Diderot, Sorbonne Paris Cit\'e, 91191, Gif-sur-Yvette, France\\
}

\date{Accepted 2021 May 8. Received 2021 May 7; in original form 2021 February 16}

\pubyear{2021}

\begin{document}
\label{firstpage}
\pagerange{\pageref{firstpage}--\pageref{lastpage}}
\maketitle

\begin{abstract}
    With recent advances in modelling stars using high-precision asteroseismology, the systematic effects associated with our assumptions of stellar helium abundance ($Y$) and the mixing-length theory parameter ($\alpha_{\rm MLT}$) are becoming more important. We apply a new method to improve the inference of stellar parameters for a sample of \emph{Kepler} dwarfs and subgiants across a narrow mass range ($0.8 < M < 1.2\,\rm M_\odot$). In this method, we include a statistical treatment of $Y$ and the $\alpha_{\rm MLT}$. We develop a hierarchical Bayesian model to encode information about the distribution of $Y$ and $\alpha_{\rm MLT}$ in the population, fitting a linear helium enrichment law including an intrinsic spread around this relation and normal distribution in $\alpha_{\rm MLT}$. We test various levels of pooling parameters, with and without solar data as a calibrator. When including the Sun as a star, we find the gradient for the enrichment law, $\Delta Y / \Delta Z = 1.05\substack{+0.28\\-0.25}$ and the mean $\alpha_{\rm MLT}$ in the population, $\mu_\alpha = 1.90\substack{+0.10\\-0.09}$. While accounting for the uncertainty in $Y$ and $\alpha_{\rm MLT}$, we are still able to report statistical uncertainties of 2.5 per cent in mass, 1.2 per cent in radius, and 12 per cent in age. Our method can also be applied to larger samples which will lead to improved constraints on both the population level inference and the star-by-star fundamental parameters.
\end{abstract}




\section{Introduction}



The inference of stellar ages, masses, and radii has improved through the use of asteroseismology in recent years \citep[e.g. see the review by][]{Chaplin.Miglio2013}. Measuring the oscillation modes in stars using photometric time series data, from missions such as \emph{CoRoT} \citep{Baglin.Auvergne.ea2006}, \emph{Kepler} \citep{Borucki.Koch.ea2010}, and \emph{TESS} \citep{Ricker.Winn.ea2015} has given us new insights into the structure and evolution of stars \citep{Garcia.Ballot2019}. Recent examples include a deeper understanding of stellar structure \citep{Verma.Raodeo.ea2017}, chronology of a Milky Way merger \citep{Chaplin.Serenelli.ea2020}, and classifying exoplanetary systems \citep{Huber.Chaplin.ea2019, Tayar.Claytor.ea2020}. Several studies used grids of stellar models with constraints from asteroseismology to produce catalogues of precise stellar parameters \citep{Pinsonneault.Elsworth.ea2014, SilvaAguirre.Lund.ea2017}. However, with increasing precision on fundamental parameters inferred from stellar models with asteroseismology, extra care should be taken to ensure that we are accounting for uncertainty in our choice of stellar physics.

Typically, stars are modelled on a star-by-star basis with estimates of stellar properties including some assumptions handled by simple scaling relations and solar calibrations. For instance, a helium ($Y$) to heavy-element ($Z$) enrichment ratio ($\Delta Y / \Delta Z$) and mixing-length theory parameter ($\mlt$) are often assumed. However, there has been little effort to account for the population distribution of such quantities. Assuming values for $Y$ and $\mlt$ can result in inaccurate inference and systematics on the inferred stellar parameters \citep{Valle.DellOmodarme.ea2015}. Independently modelling $Y$ and $\mlt$ can also be computationally demanding and requires high-precision observations in order to return meaningful stellar properties.

In this work, we apply the method of Davies et al. (in prep.) to determine stellar properties for a sample of \emph{Kepler} dwarfs and subgiants using a hierarchical Bayesian model (HBM). With an HBM, we introduce population-level distributions for $Y$ and $\mlt$ to encode prior information throughout the sample. We will show that when an HBM is used, we can increase the precision of inferred masses, radii, and ages despite increasing the number of free parameters.

The use of HBMs has been demonstrated in other areas of astrophysics to reduce individual parameter uncertainties by encoding prior information about the distribution of the parameter in a population. For example, HBMs have been used with data from \emph{Gaia}, improving distance measures \citep{Leistedt.Hogg2017, Anderson.Hogg.ea2018} and calibrating the red clump as a standard candle \citep{Hawkins.Leistedt.ea2017, Chan.Bovy2020} using asteroseismology \citep{Hall.Davies.ea2019}. In other instances, HBMs have been used to infer binary-star and exoplanet eccentricities \citep{Hogg.Myers.ea2010}, obliquity of transit systems \citep{Morton.Winn2014}, stellar inclination with asteroseismology \citep{Campante.Lund.ea2016, Kuszlewicz.Chaplin.ea2019}, and the age-metallicity relation of the solar neighbourhood \citep{Feuillet.Bovy.ea2016}.

To describe the distribution of $Y$ in this work, we assume a linear helium enrichment law characterised by freely varied population parameters: the gradient given by $\Delta Y / \Delta Z$, the primordial helium abundance at $Z=0$ ($Y_P$) and an intrinsic spread in helium ($\sigma_Y$). There have been many studies to determine a linear enrichment law, from modelling eclipsing binaries \citep{Ribas.Jordi.ea2000} to spectroscopy of galactic H\textsc{ii} regions \citep{Balser2006}. Values of $\Delta Y / \Delta Z$ have also been determined for samples of main sequence (MS) stars \citep{Casagrande.Flynn.ea2007}, open clusters \citep{Brogaard.VandenBerg.ea2012}, and more recently with asteroseismology using individual oscillation frequencies \citep{SilvaAguirre.Lund.ea2017} and the glitch due to the second helium ionisation zone \citep{Verma.Raodeo.ea2019}. In these studies the value of the enrichment ratio was typically inferred to be $1 \lesssim \Delta Y / \Delta Z \lesssim 3$. However, helium enrichment is unlikely to be exactly linear with metallicity and may depend on other chemical abundances \citep{West.Heger2013} or the location of the star in the Milky Way \citep{Frebel2010}. Therefore, we account for some deviation from the linear law by introducing $\sigma_Y$. Moreover, our method may be adapted to different helium enrichment priors in future work.

The widely used mixing-length theory (MLT) of convection, parametrised by $\mlt$, has been tested throughout the Hertzsprung-Russell (HR) diagram with 3D hydrodynamical simulations \citep{Trampedach.Stein.ea2014, Magic.Weiss.ea2015} and asteroseismology \citep{Tayar.Somers.ea2017, Viani.Basu.ea2018, Li.Bedding.ea2018} with values in the range $0.8 \lesssim \mlt/\alpha_{\mathrm{MLT}, \odot} \lesssim 1.2$ where $\alpha_{\mathrm{MLT}, \odot}$ is the value calibrated to the Sun. However, in many grids of stellar models, a constant value calibrated to reproduce the Sun is assumed. This can lead to systematic uncertainties in stellar ages due to the effects of variable mixing depending on the mixing length. The MLT approximates convective mixing and thus we expect the value of $\mlt$ to vary from star-to-star due to various effects, from changes in chemical composition to other sources of mixing described poorly by MLT. The investigation of more complex $\mlt$ distributions is beyond the scope of this work. Instead, we experiment with two prior assumptions for $\mlt$ in the population. The first assumes $\mlt$ is normally distributed in our sample with a spread ($\sigma_\alpha$) to account for the aforementioned variation in $\mlt$. The second assumes $\mlt$ is constant throughout the sample.

Our HBM requires a way to map from the stellar initial (or bulk) properties to predict observables. We can achieve this with a large grid of stellar evolutionary models. However, a discrete grid can produce inaccurate posterior distributions, limited to the grid resolution. Increasing the resolution is computationally demanding, especially when scaling to higher input dimensions. One solution is to interpolate the stellar models, as is common in the isochrone-fitting method \citep[see e.g.][]{Berger.Huber.ea2020}. However, interpolation can become computationally expensive at high input dimensions and grid size, and evaluating the likelihood using modern Bayesian sampling techniques is slow. Therefore, we use machine learning to map stellar model inputs to observables to provide a fast way to sample the HBM. In this work, we train an artificial neural network (ANN) on a large grid of stellar models. There have been similar applications of ANNs in asteroseismology \citep{Verma.Hanasoge.ea2016, Bellinger.Angelou.ea2016, Hon.Stello.ea2017, Hon.Stello.ea2018a, Hendriks.Aerts2019} but not yet in the context of an HBM. Using the machine learning speed-up, we demonstrate a scalable method for obtaining fundamental stellar parameters.

In Section \ref{sec:data}, we describe the data for the sample of 81 \emph{Kepler} dwarfs and subgiants studied in this work. We then present the methods in Section \ref{sec:meth} for which we produced a large grid of stellar evolutionary models to use as training data for an ANN. We use the ANN as an emulator in a series of statistical models described in Section \ref{sec:hbm} to model the sample and present our results in Section \ref{sec:res}. Finally, in Section \ref{sec:dis} we discuss the results from each model and compare them to results for the sample of stars in the literature.

\section{Data}\label{sec:data}



For this study, we selected the sample of 415 stars from the first APOKASC catalogue of dwarfs and subgiants \citep[][hereafter \citetalias{Serenelli.Johnson.ea2017}]{Serenelli.Johnson.ea2017}. This sample provides an extensive set of dwarfs and subgiant stars with asteroseismic detections observed by the \emph{Kepler} mission. \citetalias{Serenelli.Johnson.ea2017} used grid-based modelling to determine the ages ($\tau$), masses ($M$), radii ($R$), and surface gravity ($\log g$) of stars in the sample, using global asteroseismic parameters, effective temperature ($\teff$), and metallicity ($\metallicity$) as inputs.

Using five independent pipelines, \citetalias{Serenelli.Johnson.ea2017} determined values for global asteroseismic parameters -- the large frequency separation $\dnu$ and the frequency at maximum power, $\numax$ with median uncertainties of 1.7 per cent and 4 per cent respectively. We chose to adopt the $\dnu$ determined in their work as inputs for our method. They also used $\metallicity$ published in Data Release 13 \citep[DR13;][]{Albareti.AllendePrieto.ea2017} of the APOGEE stellar abundances pipeline \citep[ASPCAP;][]{GarciaPerez.AllendePrieto.ea2016} with uncertainties of \SI{0.1}{\dex}. For their preferred set of results, they adopted $\teff$ from the Sloan Digital Sky Survey (SDSS) \emph{griz}-band photometry \citep{Pinsonneault.An.ea2012} with a median uncertainty of \SI{70}{\kelvin}.

We removed more evolved stars from the APOKASC sample by cutting those with $\log g < \SI{3.8}{\dex}$. We then kept stars within 1-$\sigma$ of $-0.5 < \metallicity < \SI{0.5}{\dex}$ to remove metal-poor and -rich stars. Main sequence stars with $M \gtrsim \SI{1.2}{\solarmass}$ are understood to have a convective, hydrogen-burning core, with some dependence on the chemical composition and choice of stellar physics \citep{Appourchaux.Antia.ea2015}. Stellar models with a convective core require the treatment of extra stellar physics such as overshooting, which is beyond the scope of this work. Therefore, we keep only stars with masses within 1-$\sigma$ of \SIrange{0.8}{1.2}{\solarmass} from the preferred set of results of \citetalias{Serenelli.Johnson.ea2017}.

Following cuts to the sample, we adopted updated ASPCAP spectroscopic metallicities, \metallicity, from Data Release 14 \citep[DR14;][]{Blanton.Bershady.ea2017} which had a median uncertainty of \SI{0.07}{\dex}. We also chose to adopt $\teff$ from the same catalogue to be internally consistent. We note that our chosen effective temperature scale is offset from the photometric temperature scale of \citetalias{Serenelli.Johnson.ea2017} by approximately $- \SI{170}{\kelvin}$, with a dispersion of $\sim \SI{120}{\kelvin}$ corresponding to typical uncertainties on the individual temperatures. The median uncertainty in our adopted ASPCAP $\teff$ was \SI{125}{\kelvin}.

To provide a means of calculating luminosities, we used \emph{Gaia} Data Release 2 (DR2) parallaxes \citep{GaiaCollaboration.Prusti.ea2016, GaiaCollaboration.Brown.ea2018}. We cross-matched the remaining sample with the DR2 catalogue, taking the nearest neighbours within a 4 arcsec radius. Although DR2 parallaxes have improved upon the DR1 values at the time of \citetalias{Serenelli.Johnson.ea2017}, there was still evidence for a zero-point offset \citep{Lindegren.Hernandez.ea2018}. We adopted a global offset of \SI{0.05}{\milli\aarcsec}, in the sense that DR2 parallaxes were underestimated, representative of values obtained in the literature using \textit{Kepler} field stars \citep[see e.g.][]{Zinn.Pinsonneault.ea2019, Hall.Davies.ea2019} and through other methods \citep[see e.g.][]{Riess.Casertano.ea2018, Chan.Bovy2020}. We then cross-matched our sample with the Two-Micron All Sky Survey \citep[2MASS;][]{Skrutskie.Cutri.ea2006} to obtain \emph{K\textsubscript{S}}-band (\SI{2.16}{\micro\metre}) photometry.

We determined luminosities, $L$ for the sample using the direct method of \textsc{isoclassify} \citep{Huber.Zinn.ea2017, Berger.Huber.ea2020} with \emph{K\textsubscript{S}}-band photometry, \emph{Gaia} DR2 parallaxes, ASPCAP $\metallicity$ and $\teff$, and asteroseismic $\log g$ as inputs. This involved computing absolute \emph{K\textsubscript{S}}-band magnitudes using the \emph{Gaia} DR2 parallaxes and extinctions determined by the 3D galactic reddening maps of \citet{Green.Schlafly.ea2018}. We determined absolute bolometric magnitudes by interpolating the MIST bolometric correction tables as a function of $\teff$, $\log g$ and $\metallicity$ \citep{Dotter2016, Choi.Dotter.ea2016}. We adopted the uncertainty of \SI{0.02}{\magnitude} assumed by \textsc{isoclassify} for both the extinctions and bolometric corrections, corresponding to typical systematic errors in extinction maps and bolometric fluxes \citep[e.g.][]{Zinn.Pinsonneault.ea2019a,Tayar.Claytor.ea2020}. We obtained luminosities for the sample with a median uncertainty of 3.4 per cent.

\begin{figure}
    \centering
    \includegraphics[width=\linewidth]{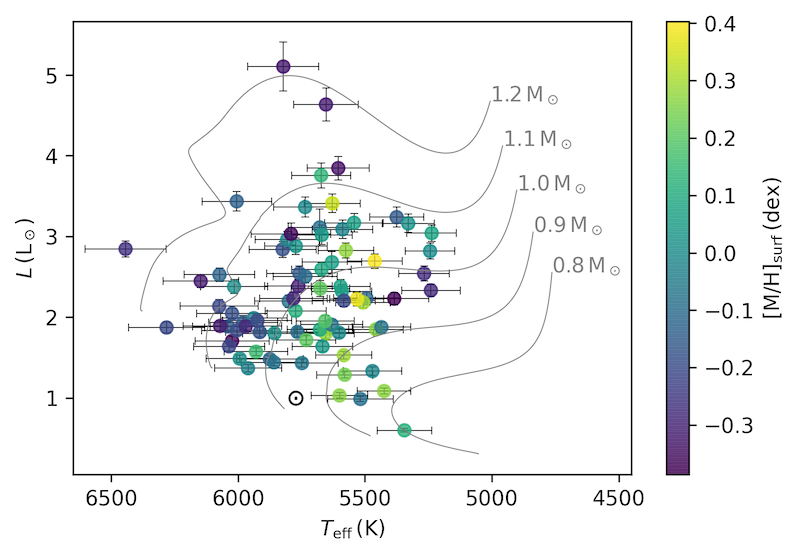}
    \caption{The luminosity, $L$ against effective temperature, $\teff$ of the sample of 81 \emph{Kepler} dwarfs and subgiants studied in this work. Each stars is coloured according to metallicity. The grey lines depict evolutionary tracks with $\metallicity_\mathrm{init}=\SI{0.0}{\dex}$, $Y_\mathrm{init}=0.28$ and $\mlt=1.9$ for different stellar masses. The current position of the Sun is shown by the $\odot$ symbol.}
    \label{fig:data}
\end{figure}

The final sample comprised 81 stars for which we had complete data for $\teff$, $\metallicity$, $\dnu$, and $L$ to use as inputs for our stellar modelling method -- see Table \ref{tab:data}. In Fig. \ref{fig:data}, we show the HR diagram for the sample plot in context with a series of stellar evolutionary tracks at solar metallicity.

\begin{table*}
	\centering
	\caption{The observables and their respective uncertainties for 5 stars in the sample of 81 stars. The whole table is available as supplementary material.}
	\label{tab:data}
	\input{tables/stars_inputs.tex}
\end{table*}

\section{Methods}\label{sec:meth}



Firstly, we used a stellar evolutionary code to compute a grid of models to predict observable quantities (see Section \ref{sec:grid}). Subsequently, we trained an ANN on the grid of stellar models to map input parameters to output observables (see Appendix \ref{sec:ann} for further details). We then constructed three Bayesian models in Section \ref{sec:hbm} which each sampled the trained ANN to estimate stellar fundamental parameters as described in Section \ref{sec:sampling}. Evaluation of the ANN gradient is required during training. Consequently, estimating the gradient of the model likelihood is fast and simple when the observables are generated by an ANN. Hence, we open up the possibility of using a Hamiltonian Monte Carlo (HMC) algorithm -- for example, using the No-U-Turn Sampler \citep[NUTS;][]{Hoffman.Gelman2014} -- which requires the gradient to sample the model posterior. Once we had tested the accuracy of the model on a sample of synthetic stars, we evaluated each model on the subset of the APOKASC catalogue selected in Section \ref{sec:data}.

\subsection{Grid of stellar models}\label{sec:grid}



We built a stellar model grid to use in training the ANN. The grid includes four independent model inputs: stellar mass ($M$), initial helium fraction ($Y_{\rm init}$), initial metallicity ($\mathrm{[M/H]}_\mathrm{init}$), and the mixing-length parameter ($\mlt$). Ranges and grid steps of the four model inputs are summarised in Table \ref{tab:grid}. We increased the resolution at higher $\metallicity$ to give a more consistent resolution in $Z_\mathrm{init}$ which we used as an ANN input in Appendix \ref{sec:ann}. We computed each stellar evolutionary track from the Hayashi line to the base of red-giant branch defined here by $\log g$ = 3.6 dex. 
We also computed an additional set of evolutionary tracks, with input values chosen randomly within the range of the regular grid, to use when testing the accuracy of the ANN.

\begin{table}
	\centering
	\caption{Stellar model grid parameters for train and test datasets. $N_\mathrm{track}$ are the numbers of stellar evolutionary tracks for each dimension of the grid, multiplied to a total of \num{17220} tracks.}
	\label{tab:grid}
	\begin{tabular}{cccc} 
		\toprule
		\multicolumn{4}{c}{\textbf{Stellar model grid}}\\
		Input Parameter & Range & Increment & $N_{\rm track}$\\
        \midrule
	$M$ ($M_{\odot}$)  & 0.80 -- 1.20 &  0.01& \num{41}\\
        $\rm{[M/H]}_\mathrm{init}$ (dex) & -0.5 -- 0.2/0.25 -- 0.5 & 0.1/0.05 & \num{14}\\
        	$Y_{\rm init}$ & 0.24 -- 0.32 & 0.02 & \num{5}\\
        $\alpha_{\rm{mlt}}$  & 1.5 -- 2.5&  0.2 & \num{6} \vspace{0.5em}\\
        \textbf{Total} & & & \num{17220}\\
        \bottomrule
    \end{tabular}
\end{table}

\subsubsection{Stellar models and input physics}\label{subsec:stellar_model}

We used Modules for Experiments in Stellar Astrophysics
(\textsc{MESA}, version 12115) to establish a grid of stellar models. 
\textsc{MESA} is an open-source stellar evolution package which is undergoing active development. 
Descriptions of input physics and numerical methods
can be found in \citet{Paxton.Bildsten.ea2011, Paxton.Cantiello.ea2013, Paxton.Marchant.ea2015, Paxton.Schwab.ea2018, Paxton.Smolec.ea2019}.
We adopted the solar chemical mixture, $(Z/X)_{\odot}$ = 0.0181,
 provided by \citet{Asplund.Grevesse.ea2009}. 
The initial chemical composition was calculated by:
\begin{equation}
\log (Z_{\rm{init}}/X_{\rm{init}}) = \log (Z/X)_{\odot} + \rm{[M/H]_{init}}.  \\
\end{equation}
We used the \textsc{MESA} $\rho-T$ tables based on the 2005
update of OPAL EOS tables \citep{Rogers.Nayfonov2002} and OPAL opacity
supplemented by low-temperature opacity \citep{Ferguson.Alexander.ea2005}. The grey Eddington $T-\tau$ relation is used to determine boundary conditions for modelling the atmosphere. The MLT of convection was implemented, where 
$\alpha_{\rm MLT} = \ell_{\rm MLT}/H_p$ is the mixing-length parameter. 
We also applied the \textsc{MESA} predictive mixing scheme \citep{Paxton.Schwab.ea2018, Paxton.Smolec.ea2019} in the model computation. 

Atomic diffusion of helium and heavy elements was also taken into account. \textsc{MESA} calculates particle diffusion and gravitational settling by solving Burger's equations using the method
and diffusion coefficients of \citet{Thoul.Bahcall.ea1994}. We considered eight elements (${}^1{\rm H}, {}^3{\rm He}, {}^4{\rm He}, {}^{12}{\rm C}, {}^{14}{\rm N}, {}^{16}{\rm O}, {}^{20}{\rm Ne}$, and ${}^{24}{\rm Mg}$)
for diffusion calculations, and had the charge calculated by the \textsc{MESA} ionization module, which estimates the typical ionic charge as a function of $T$, $\rho$, and free electrons per nucleon from \citet{Paquette.Pelletier.ea1986}.

\subsubsection{Oscillation models and asteroseismic quantities}\label{subsec:seismo_model}

Theoretical stellar oscillations were calculated with the \textsc{GYRE} code (version 5.1), which was developed by \citet{Townsend.Teitler2013}. We computed radial modes (for $\ell$ = 0) for 42 radial orders by solving the adiabatic stellar pulsation equations with the structural models generated by \textsc{MESA}. We determined the asteroseismic large separation ($\dnu$) for each model with theoretical radial modes to avoid the systematic offset of the scaling relation. We derived $\Delta \nu$ with the approach given by \citet{White.Bedding.ea2011}, which is a weighted least-squares fit to the radial frequencies as a function of $n$.

We chose to ignore the well known, yet poorly characterised impact of modelled oscillation mode inaccuracies in the near-surface region of the star \citep{Kjeldsen.Bedding.ea2008, Ball.Gizon2014, Sonoi.Samadi.ea2015}. This typically presents only a small effect compared to observational uncertainties when considering the average large frequency spacing, $\dnu$. Additionally, there may be further inaccuracies in the modelled $\dnu$ because of variations in p mode frequencies with stellar activity \citep{Chaplin.Elsworth.ea2007, Garcia.Mathur.ea2010, Kiefer.Schad.ea2017}. Therefore, a thorough treatment of systematic uncertainties in $\dnu$ is instead left to future work (Carboneau et al. in preparation). 

\subsection{Statistical models}\label{sec:hbm}



We devised three Bayesian models, each with varying levels of parameter sharing (pooling) between stars in the population. Initially, we tested the models and demonstrated reduction of statistical uncertainties in the stellar fundamental parameters by analysing a random sample of 100 synthetic stars modelled using \textsc{MESA}. Then, we applied the models to the sample of stars in Table \ref{tab:data} (with and without solar data for two of the models) and compared the results with that of \citetalias{Serenelli.Johnson.ea2017}.

Our first model was equivalent to modelling each star individually and featured no pooling; henceforth, we refer to it as the no-pooled (NP) model (see Section \ref{sec:np}). We then derived two hierarchical Bayesian models (HBMs) which use population-level parameters to describe their distribution in the sample. Both of these models partially-pooled helium using a linear enrichment law. We drew the initial helium fraction for each star from a normal distribution with a mean described by the enrichment law and standard deviation representing its spread. Similarly, we partially-pooled the MLT parameter, $\mlt$ in one model, whereas we maximally-pooled $\mlt$ in the other, such that it assumes the same value for the entire sample. Hence, we refer to the former as the partial-pooled (PP) model and the latter as the max-pooled (MP) model, described in Sections \ref{sec:pp} and \ref{sec:mp} respectively.

\subsubsection{No-pooled model}\label{sec:np}

Firstly, we constructed a model comprising independent parameters corresponding to the ANN inputs $\boldsymbol{\theta}_i = \{f_{\mathrm{evol}, i}, M_i, \alpha_{\mathrm{MLT},i}, Y_i, Z_i\}$ for a given star, $i$. The parameter $f_{\mathrm{evol}, i}$ describes the evolutionary stage of the $i$-th star (see Equation \ref{eq:fevol} in the Appendix). Using Bayes' theorem, the \emph{posterior} probability density function (PDF) of the model parameters given a set of observed data, $\boldsymbol{d}_i$ is,
\begin{equation}
    p(\boldsymbol{\theta}_i | \boldsymbol{d}_i) \propto p(\boldsymbol{\theta}_i) \, p(\boldsymbol{d}_i | \boldsymbol{\theta}_i)\,,
    \label{eq:bayes}
\end{equation}
where $p(\boldsymbol{\theta}_i)$ is the \emph{prior} PDF of the model parameters and $p(\boldsymbol{d}_i | \boldsymbol{\theta}_i)$ is the \emph{likelihood} of observing the data given the model.

We chose weakly-informative, bounded priors for the independent parameters, restricting them to their respective ranges in the ANN training data. Although the neural network is able to make predictions outside the training data range, these have not been tested and may be unreliable. Therefore, we used a beta distribution with $\alpha = \beta = 1.2$ as the prior PDF on the independent parameters, transformed such that the probability is null outside the chosen range,
\begin{equation}
    p(\boldsymbol{\theta}_i) \propto \prod_{k=1}^{N_{\theta}} \mathcal{B}\left(\tilde{\theta}_{k, i} | 1.2, 1.2\right),
\end{equation}
where the beta distribution is defined as,
\begin{equation}
    \mathcal{B}(x | \alpha, \beta) = \frac{x^{\,\alpha-1}(1-x)^{\,\beta-1}}{\int_{0}^{1} u^{\,\alpha-1}(1-u)^{\,\beta-1} \mathrm{d} u}.
    \label{eq:beta}
\end{equation}
and,
\begin{equation}
    \tilde{\theta}_{k, i} = \frac{\theta_{k, i} - \theta_{k, \mathrm{min}}}{\theta_{k, \mathrm{max}} - \theta_{k, \mathrm{min}}},
\end{equation}
is the transformed parameter where $\theta_{k, \mathrm{min}}$ and $\theta_{k, \mathrm{max}}$ are the upper and lower bounds for each parameter. The beta distribution was preferred over a bounded uniform distribution because our sampler evaluates the gradient of the posterior and is thus sensitive to discontinuities (see e.g. Fig. \ref{fig:beta} in the Appendix).

Using notation which represents a given random variable $x \sim q$ as equivalent to being drawn from a probability distribution $p(x) \propto q(x)$ where $q(x)$ is a non-normalised probability density function, we may write the priors for $\boldsymbol{\theta}_i$ as,
\begin{align*}
    f_{\mathrm{evol}, i} &\sim 0.01 + 1.99 \cdot \mathcal{B}(1.2, 1.2),\\
    M_i &\sim 0.8 + 0.4 \cdot \mathcal{B}(1.2, 1.2),\\
    \alpha_{\mathrm{MLT}, i} &\sim 1.5 + \mathcal{B}(1.2, 1.2),\\
    Y_{\mathrm{init}, i} &\sim 0.24 + 0.08 \cdot \mathcal{B}(1.2, 1.2),\\
    Z_{\mathrm{init}, i} &\sim 0.005 + 0.035 \cdot \mathcal{B}(1.2, 1.2),\\
\end{align*}
where each beta distribution is scaled to cover the boundaries of the grid of stellar models computed in Section \ref{sec:grid}.

We made predictions for each star using the trained ANN, $\{\log(\tau)_i, T_{\mathrm{eff}, i}, R_i, \dnu_i, \metallicity_{\mathrm{surf}, i}\} = \boldsymbol{f}_{\mathrm{ANN}}(\boldsymbol{\theta}_i)$, from which we derived the luminosity, $L_i$ using the Stefan-Boltzmann law. Out of the model parameters, those which may be observed are denoted by ${\boldsymbol{\mu}}_{i} = {\boldsymbol{f}}(\boldsymbol{\theta}_i)$. Therefore, we write the likelihood that we observe any $\boldsymbol{d}_i$ with known uncertainty, $\boldsymbol{\sigma}_{i}$ given our model as,
\begin{equation}
    p(\boldsymbol{d}_i | \boldsymbol{\theta}_i) = \prod_{k=1}^{N_\mathrm{obs}} \frac{1}{\sigma_{k, i} \sqrt{2\pi}} \exp\left[ - \frac{(d_{k, i} - \mu_{k, i})^2}{2 \sigma_{k, i}^2} \right],
    \label{eq:like}
\end{equation}
where $N_\mathrm{obs}$ is the number of observed variables. We chose to use observed $\teff$, $L$, $\dnu$, and $\metallicity$ collated for our sample as described in Section \ref{sec:data}.

It follows that the posterior PDF for a population of $N_\mathrm{stars}$ stars for the NP model is, 
\begin{equation}
    p(\boldsymbol{\Theta} | \boldsymbol{D}) = \prod_{i=1}^{N_{\mathrm{stars}}} p(\boldsymbol{\theta}_i | \boldsymbol{d}_i),   
\end{equation}
where $\boldsymbol{\Theta}$ is the matrix of model parameters and $\boldsymbol{D}$ is the matrix of observables. A probabilistic graphical model (PGM) of the NP model can be seen inside the grey box of Fig. \ref{fig:pgm}, without the arrow connecting $Z_\mathrm{init}$ to $Y_\mathrm{init}$. We ignore the nodes outside the box because these correspond the the PP model described next.

\subsubsection{Partial-pooled model}\label{sec:pp}



Sharing, or pooling parameters between stars in a population can improve the uncertainties on stellar fundamentals by encoding our prior knowledge of their distribution in a population. We constructed a hierarchical model, which builds upon the NP model by introducing population-level \emph{hyperparameters}. Specifically, we chose to describe initial helium and $\mlt$ by partially-pooling them.

We constructed the PP model such that each of the initial helium, $\boldsymbol{Y}_\mathrm{init}$ and MLT parameter, $\boldsymbol{\alpha}_\mathrm{MLT}$ are drawn from a common distribution characterised by the set of hyperparameters, $\boldsymbol{\phi}$. Thus, Bayes' theorem becomes,
\begin{equation}
    p(\boldsymbol{\phi}, \boldsymbol{\Theta} | \boldsymbol{D}) \propto p(\boldsymbol{\phi}) \, p(\boldsymbol{Y}_\mathrm{init}, \boldsymbol{\alpha}_\mathrm{MLT} | \boldsymbol{\phi}) \, p(\boldsymbol{f}_{\mathrm{evol}}, \boldsymbol{M}, \boldsymbol{Z}) \, p(\boldsymbol{D} | \boldsymbol{\Theta}),
    \label{eq:hbmbayes}
\end{equation}
where $\boldsymbol{\Theta}$ is the same as in the NP model, i.e. each object-level parameter, $\boldsymbol{\theta}_j = \{\theta_{j, i}\}_{i=1}^{N_\mathrm{stars}}$, and $\boldsymbol{\phi} = \{\Delta Y/\Delta Z, Y_P, \sigma_Y, \mu_\alpha, \sigma_\alpha\}$. The hyperparameters for $\boldsymbol{Y}_\mathrm{init}$ comprise the helium enrichment ratio (${\Delta Y}/{\Delta Z}$), primordial helium abundance fraction ($Y_P$), and the spread in helium ($\sigma_Y$). The remaining hyperparameters for $\boldsymbol{\alpha}_\mathrm{MLT}$ comprise the mean ($\mu_\alpha$) and spread, ($\sigma_\alpha$).

We assumed the initial helium and the mixing-length parameter are each drawn from a normal distribution characterised by a population mean and standard deviation. The probability of $\boldsymbol{Y}_\mathrm{init}$ and $\boldsymbol{\alpha}_\mathrm{MLT}$ given $\boldsymbol{\phi}$ is,
\begin{equation}
    p(\boldsymbol{Y}_\mathrm{init}, \boldsymbol{\alpha}_\mathrm{MLT} | \boldsymbol{\phi}) = p(\boldsymbol{Y}_\mathrm{init} | \boldsymbol{\mu}_Y, \sigma_Y) \, p(\boldsymbol{\alpha}_\mathrm{MLT} | \mu_\alpha, \sigma_\alpha),
    \label{eq:ppool}
\end{equation}
where $\boldsymbol{\mu}_Y$ and is the mean initial helium fraction as described by the linear helium enrichment law,
\begin{equation}
    \boldsymbol{\mu}_{Y} = Y_P + \frac{\Delta Y}{\Delta Z} \boldsymbol{Z}_{\mathrm{init}}.\label{eq:helium}
\end{equation}
Therefore, we may write the prior PDF of initial helium given its population-level hyperparameters as,
\begin{equation}
    p(\boldsymbol{Y}_{\mathrm{init}} | \boldsymbol{Z}_{\mathrm{init}}, {\Delta Y}/{\Delta Z}, Y_P, \sigma_Y) = \prod_{i=1}^{N_\mathrm{stars}} \mathcal{N}({Y}_{\mathrm{init}, i} | {\mu}_{Y, i}, \sigma_Y).
\end{equation}

Similarly, for the second component of Equation \ref{eq:ppool}, we chose to partially-pool $\mlt$. We assume that convection in stars of a similar mass, evolutionary stage and area of the HR diagram may be approximated using a similar value of $\mlt$, but the accuracy of the MLT may vary from star-to-star. Given the small range of our sample, any such variation will be absorbed by the spread parameter, $\sigma_\alpha$. Therefore, we decided to describe the prior on $\boldsymbol{\alpha}_\mathrm{MLT}$ as,
\begin{equation}
    p(\boldsymbol{\alpha}_{\mathrm{MLT}} | \mu_\alpha, \sigma_\alpha) = \prod_{i=1}^{N_\mathrm{stars}} \mathcal{N}({\alpha}_{\mathrm{MLT}, i} | \mu_\alpha, \sigma_\alpha).
\end{equation}

We gave all of the hyperparameters weakly informative priors, with the exception of $Y_P$ for which we adopt a recent measurement of the primordial helium abundance from big bang nucleosynthesis (BBN) as the mean \citep{Pitrou.Coc.ea2018}, with a standard deviation representative of the range of values in the literature \citep{Aver.Olive.ea2015, Peimbert.Peimbert.ea2016, Cooke.Fumagalli2018}. Hence, we assumed priors on the hyperparameters as follows,
\begin{align*}
    {\Delta Y}/{\Delta Z} &\sim 4.0\cdot\mathcal{B}(1.2, 1.2),\\
    Y_P &\sim \mathcal{N}(0.247, 0.001),\\
    \sigma_Y &\sim \ln\mathcal{N}(0.01, 1.0),\\
    \mu_\alpha &\sim 1.5 + \mathcal{B}(1.2, 1.2),\\
    \sigma_\alpha &\sim \ln\mathcal{N}(0.1, 1.0),
\end{align*}
where, $x \sim \ln\mathcal{N}(m, \sigma)$ represents a random variable drawn from the log-normal distribution,
\begin{equation}
    \ln\mathcal{N}(x | m, \sigma)=  \frac{1}{x \sigma \sqrt{2 \pi}} \exp \left[ - \frac{\ln (x / m)^{2}}{2 \sigma^{2}}\right].
\end{equation}

\begin{figure}
    \includegraphics[width=\linewidth]{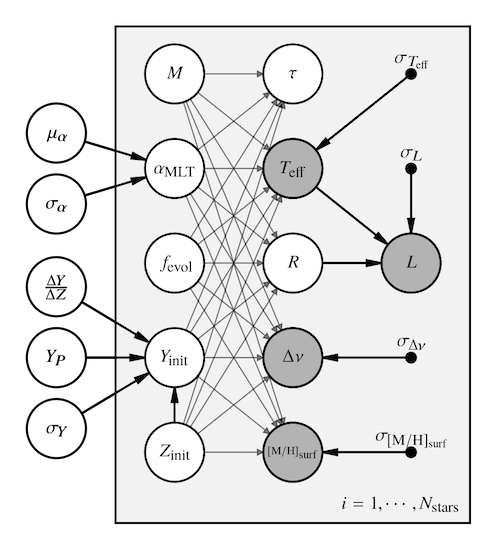}
    \caption{A probabilistic graphical model (PGM) of the partially-pooled (PP) hierarchical model. Nodes outside of the grey rectangle represent the hyperparameters in the model. Nodes inside the grey rectangle represent individual stellar parameters. Dark grey nodes represent observables which each have their respective observational uncertainties given by the solid black nodes. The direction of the arrows represent the dependencies in the generative model.}
    \label{fig:pgm}
\end{figure}

We produced a PGM for the model, depicted in Fig. \ref{fig:pgm}. The hyperparameters are shown outside of the grey box containing the individual stellar parameters to represent the hierarchical aspect of the model.

\subsubsection{Max-pooled model}\label{sec:mp}



We built another hierarchical model similar to the PP model except that $\mlt$ is max-pooled (MP). In this model, we assumed that $\mlt$ must be the same value for every star in the sample, but still allowed it to freely vary on a population-level. Thus the hyperparameters are now, $\boldsymbol{\phi} = \{\Delta Y/\Delta Z, Y_P, \sigma_Y, \mlt\}$. The posterior distribution of the model takes the same form as in Equation \ref{eq:hbmbayes} except that the MLT parameter for the $i$-th star is,
\begin{equation}
    \alpha_{\mathrm{MLT}, i} = \mlt,
\end{equation}
where,
\begin{equation}
    \mlt \sim 1.5 + \mathcal{B}(1.2, 1.2),
\end{equation}
chosen such that $\mlt$ is confined to the boundaries of the grid of stellar models ($1.5 < \mlt < 2.5$).

\subsection{The Sun as a star}\label{sec:sun}

Pooling parameters in an HBM allows us to use the Sun as a calibrator in a unique way. Rather than calibrating our model physics to the Sun and then assuming the calibrated parameters across our sample, we can include the Sun as a part of the same population as our sample of stars. If we assume $Y_\mathrm{init}$ and $\mlt$ for the Sun are a part of the same prior distribution as for the rest of the sample, then we can simply add solar observables to our model inputs.

For both the PP and MP models, we iterated with and without data for the Sun included in the population, referred to as PPS and MPS respectively. We adopted the solar data in Table \ref{tab:sun} with uncertainties conservatively limited to the accuracy of the ANN for $R$, $L$ and representative of variation in the literature for $\teff$. We also adopted $\dnu=\SI{135.1(2)}{\micro\hertz}$ with a central value from \citet{Huber.Bedding.ea2011} and a standard deviation representative of variations in measurements of the solar $\dnu$ \citep{Broomhall.Chaplin.ea2011}.

\begin{table}
    \centering
    \caption{Solar input data. The references correspond to the central values and the uncertainties are chosen to either be representative of the ANN accuracy or the spread of values in the literature (see text for details).}
    \label{tab:sun}
    \input{tables/sun_inputs.tex}
\end{table}

\subsection{Sampling}\label{sec:sampling}

We obtained results for each of the models described above by sampling their posterior distributions using a Markov chain Monte Carlo (MCMC) algorithm. In particular, we used the NUTS algorithm implemented in \textsc{TensorFlow Probability} \citep[\textsc{TFP};][]{Abadi.Barham.ea2016, Dillon.Langmore.ea2017}\footnote{We interacted with \textsc{TFP} using the now deprecated \textsc{PyMC4} package, developed as a successor to \textsc{PyMC3} \citep{Salvatier.Wiecki.ea2016}.}. For each model, we produced \num{20000} samples split across \num{10} MCMC chains and computed summary statistics for the marginalised posteriors of each parameter in the model. We removed stars with problems during tuning using the Gelman-Rubin diagnostic \citep[$\hat{r}$;][]{Gelman.Rubin1992}. We used results from each model once $\hat{r} < 1.04$ for all parameters, indicating good model convergence.

Initially, we created a random synthetic population of stars using \textsc{MESA} to test the ability of the method to recover stellar properties according to our choice of model physics and population priors. We tested the NP, PP, and MP models. Since our sample was fictitious, it would not have been appropriate to include real solar data. We summarise the results for the synthetic stellar parameters and hyperparameters in Appendix \ref{sec:test-stars}. We found that the models were able to recover the true synthetic properties accurately, with increased precision when pooling parameters.

Once we had tested the method on synthetic stars, we obtained results for the sample of 81 dwarfs and subgiants described in Section \ref{sec:data}. Here, we included the PPS and MPS to test the effects of adding the Sun as a star in our population. For the purpose of comparison, we fit the hyperparameters of the PP model ($\Delta Y/\Delta Z, Y_P, \sigma_Y, \mu_\alpha, \sigma_\alpha$) to the results from the NP model.

Since our initial sample was chosen based on masses from \citetalias{Serenelli.Johnson.ea2017}, we expected some stars to lie outside (or near the boundary) of the observational parameter space provided by our grid of stellar models. We used an initial run of the NP model to catch and remove these stars. During the initial run, we dropped 16 of the 81 stars from the sample. Of the removed stars, we found the posteriors in $M$ for 6 skewed towards the prior upper mass limit of \SI{1.2}{\solarmass}. The remaining 10 removed stars suffered poor convergence during sampling ($\hat{r} >> 1.04$) which could be because of poor step-size tuning and sampling at the prior boundary.

Out of the remaining 65 stars with results from the NP model, 2 stars were dropped from the PP model. A consequence of partially pooling parameters is that a population spread, $\sigma$ allows for individual parameters to vary outside of the prior range given to the population mean, $\mu$. In this case, individual stellar $\mlt$ was allowed to vary outside the range for which the ANN was valid if $\sigma_\alpha$ was large. The 2 removed stars happened to have high likelihoods outside of the valid $\mlt$ range. We found that removing the same 2 stars from the other models made negligible difference to the results, so we leave a solution to this problem to future work. Naturally, we did not see the same issue in the MP model, so we proceeded with modelling the same 65 stars as with the NP model.

\section{Results}\label{sec:res}



In this section, we present the results for each of the NP, PP, and MP models with the sample of 81 APOKASC dwarfs and subgiants as inputs. We also present the results for the PPS and MPS models which include the Sun as a star in the population. Firstly, we show the reduction in age, mass and radius uncertainty with the addition of pooling in Section \ref{sec:param-results}. We then show the results for model hyperparameters in Section \ref{sec:hparam-results} where we infer the initial helium abundance and mixing-length parameter distribution in the sample.

\subsection{Stellar parameter results}\label{sec:param-results}

In Table \ref{tab:np}, we present results for the 65 APOKASC stars from the NP model. Running the NP model with synthetic stars resulted in unreliable uncertainties (see Appendix \ref{sec:test-stars}). This was because the boundary of the priors in $Y_\mathrm{init}$ and $\mlt$ truncated the posterior distribution leading to underestimated uncertainties and skewing their posterior means towards the centre of their priors. Therefore, we present the NP results only for comparison purposes, but we exclude them from further discussion. In Tables \ref{tab:pp} and \ref{tab:pps} we present the results for the 63 stars from the PP and PPS model respectively. In Tables \ref{tab:mp} and \ref{tab:mps} we also present results for the 65 stars from the MP and MPS models respectively. We note that for the MP models, there is no column for $\mlt$ because this parameter is the same across the population and hence is given in Section \ref{sec:hparam-results}.

\begin{table*}
	\centering
	\caption{The median of the marginalised posterior samples for each parameter output by the NP model, with their respective upper and lower 68 per cent credible intervals. The full table is available as supplementary material.}
	\label{tab:np}
	\input{tables/stars_outputs_NP.tex}
\end{table*}

\begin{table*}
	\centering
	\caption{The same as Table \ref{tab:np}, but for the PP model.}
	\label{tab:pp}
	\input{tables/stars_outputs_PP.tex}
\end{table*}

\begin{table*}
	\centering
	\caption{The same as Table \ref{tab:np}, but for the PPS model.}
	\label{tab:pps}
	\input{tables/stars_outputs_PPS.tex}
\end{table*}

\begin{table*}
	\centering
	\caption{The same as Table \ref{tab:np}, but for the MP model.}
	\label{tab:mp}
	\input{tables/stars_outputs_MP.tex}
\end{table*}

\begin{table*}
	\centering
	\caption{The same as Table \ref{tab:np}, but for the MPS model.}
	\label{tab:mps}
	\input{tables/stars_outputs_MPS.tex}
\end{table*}

\subsection{Population parameter results}\label{sec:hparam-results}

We obtained values for the hyperparameters for each of the models and present them in Table \ref{tab:hparam_results} along with their upper and lower 68 per cent credible regions. We omit the results for $Y_P$ because its posterior is the same as the prior, $Y_P=0.247\pm0.001$ for all the models. We fit the same hyperparameters from the PP model to the NP model results for $Y_\mathrm{init}$, $Z_\mathrm{init}$, and $\mlt$ for the purpose of comparison. However, the NP model results suffer from boundary effects which makes the resulting fit unreliable, pushing the population mean to the centre of the priors and underestimating the uncertainties. We leave the NP results here only for completeness.

\begin{table*}
	\centering
	\caption{Hyperparameter results for each model with the omission of $Y_P$.}
	\label{tab:hparam_results}
	\input{tables/hyperparam_results.tex}
\end{table*}

\begin{figure*}
    \begin{subfigure}[b]{.5\linewidth}
        \centering
        \includegraphics[width=\textwidth]{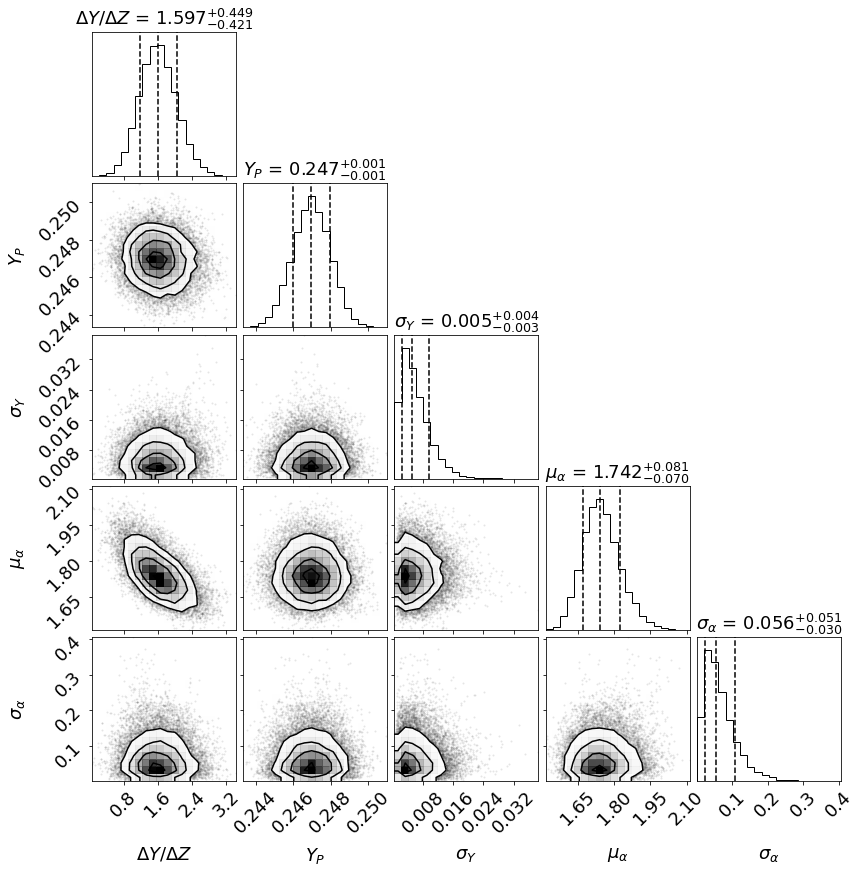}
    \end{subfigure}%
    \begin{subfigure}[b]{.5\linewidth}
        \centering
        \includegraphics[width=\textwidth]{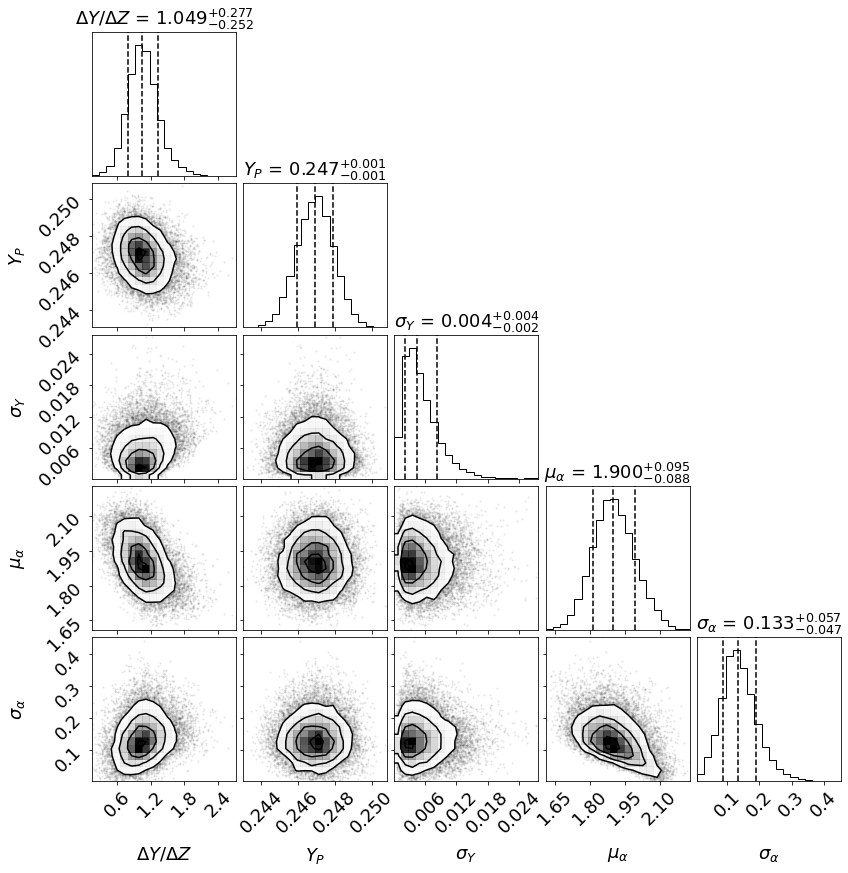}
    \end{subfigure}
    \caption{Corner plots showing the joint and marginalised sampled posterior distributions for the hyperparameters for the PP (left) and PPS (right) models. The vertical dashed lines give the 16th, 50th and 84th percentiles.}
    \label{fig:corners-pp}
\end{figure*}

\begin{figure*}
    \begin{subfigure}[b]{.5\linewidth}
        \centering
        \includegraphics[width=\textwidth]{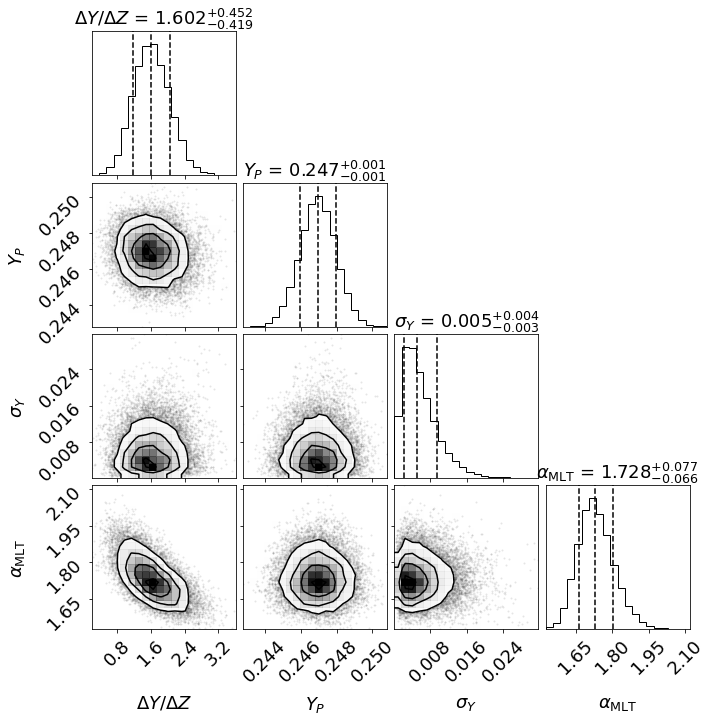}
    \end{subfigure}%
    \begin{subfigure}[b]{.5\linewidth}
        \centering
        \includegraphics[width=\textwidth]{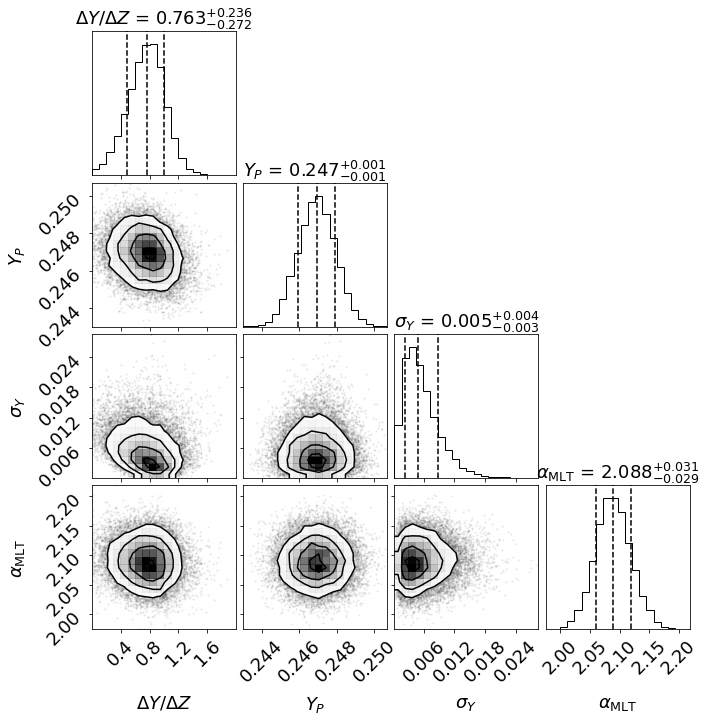}
    \end{subfigure}
    \caption{The same as Fig. \ref{fig:corners-pp} but for the MP (left) and MPS (right) models.}
    \label{fig:corners-mp}
\end{figure*}

Fig. \ref{fig:corners-pp} shows the joint and marginal distributions (corner plot) output by the PP and PPS model. We see an anti-correlation between $\Delta Y / \Delta Z$ and $\mu_\alpha$, expected due to the degeneracy between the two parameters in the stellar evolutionary models. In Fig. \ref{fig:corners-mp}, we also show the corner plot for the MP and MPS model output. Similarly, we see an anti-correlation between $\Delta Y / \Delta Z$ and $\mlt$.

\begin{figure}
    \centering
    \includegraphics[width=\linewidth]{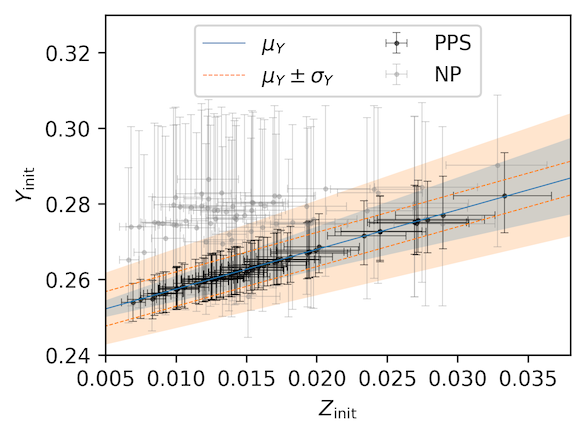}
    \caption{The results for initial helium fraction ($Y_\mathrm{init}$) against initial heavy-element fraction ($Z_\mathrm{init}$) for each star from the PPS model are shown by the black markers. The mean helium enrichment, $\mu_Y = Y_P + (\Delta Y / \Delta Z) Z_\mathrm{init}$ with its 68 per cent credible interval are shown in blue by a solid line and shaded region respectively. The population spread, $\mu_Y \pm \sigma_Y$ and its 68 per cent credible interval are shown in orange by a dashed line and shaded region respectively. The individual results from the NP model are shown by the light grey markers.}
    \label{fig:helium}
\end{figure}

We present the helium enrichment relation resulting from the PPS model in Fig. \ref{fig:helium}. In this figure, we plot the individual results for $Y_\mathrm{init}$ and $Z_\mathrm{init}$ for each of the stars from the NP and PPS models. This is an example of shrinkage in the HBM; the estimates for individual stellar parameters move towards the mean of the population when they are pooled.

\section{Discussion}\label{sec:dis}



So far, we have shown that we can add parameters to stellar models without sacrificing statistical uncertainties through the application of an HBM. We freed the $Y_\mathrm{init}$ and $\mlt$ using pooling to encode our prior knowledge of their distribution in the population. We also tested the impact of including the Sun as a star in our population. We first discuss the impact of pooling and our choice of population priors for $Y_\mathrm{init}$ and $\mlt$ in Section \ref{sec:helium} and \ref{sec:mlt}. To assess the accuracy of our model with respect to the literature, we compare our results to those of \citetalias{Serenelli.Johnson.ea2017} in Section \ref{sec:comp}. We found good agreement between this work and their results, despite some differences in observables and stellar model physics which we discuss further. Then, we discuss sources of systematic uncertainties in Section \ref{sec:sys}. Although we have accounted for uncertainties in \ref{sec:helium} and \ref{sec:mlt} in our model, there are still differences between stellar modelling codes and other model physics which should be considered. Finally, in Section \ref{sec:out}, we highlight a possible outlier in our dataset.

\subsection{Helium enrichment}\label{sec:helium}

We found the value for the helium enrichment ratio, $\Delta Y / \Delta Z$ to be the same in both the PP and MP models, $\Delta Y / \Delta Z = 1.6\substack{+0.5\\-0.4}$. This is consistent with values of $\sim 1.4$ in the literature (when heavy element diffusion is included) albeit obtained through different methods: e.g. measuring the metallicity and helium abundance of an open cluster and comparing with the primordial helium abundance \citep{Brogaard.VandenBerg.ea2012}, and fitting to helium abundances determined for \emph{Kepler} field stars using asteroseismology \citep{Verma.Raodeo.ea2019}.

When we added the Sun to the pooled models, PPS and MPS, obtained $\Delta Y / \Delta Z$ up to 2-$\sigma$ lower than the models without the Sun. In both models, the resulting $\Delta Y / \Delta Z$ of approximately \numrange{0.8}{1.0} was consistent with the initial helium fraction expected from solar models with our choice of \citet{Asplund.Grevesse.ea2009} abundances \citep{Serenelli.Basu2010}. However, such solar models have been shown to not recover helioseismic measurements of helium in the Sun \citep{Basu.Antia2004, Serenelli.Basu.ea2009, Villante.Serenelli.ea2014}. Solar models with the older \citet{Grevesse.Sauval1998} abundances typically yield higher helium fractions more in-line with helioseismology. The $\Delta Y / \Delta Z$ from the PP and MP models are higher than those including the Sun. We could extend our model to include asteroseismic indicators of helium to improve the uncertainties on $Y$ and test whether this difference becomes more significant.


Our models assumed a prior of $Y_P = \num{0.247(1)}$ for the primordial helium fraction which dominated its posterior. This was a sensible assumption to make when using a linear enrichment law, because measurements of the primordial helium correspond to the abundance at the epoch of BBN according to current cosmological theory \citep{Cyburt.Fields.ea2016}. However, if we used a less informative prior for $Y_P$ we might yield more uncertain results, or even a different value for $Y_P$. In previous work fitting a linear enrichment law, some results for $Y_P$ suggested a value below the BBN value \citep{Casagrande.Flynn.ea2007, SilvaAguirre.Lund.ea2017}. It is more probable that the assumption of a linear enrichment law is inaccurate than a sample of stars could contradict independent $Y_P$ from cosmology. We justify our prior on $Y_P$ as in-line with the assumption of a linear enrichment law, but highlight the need to investigate other ways of describing helium in a population of stars.

\subsection{Mixing-length theory}\label{sec:mlt}

To a greater degree than chemical composition, the best-fitting $\mlt$ depends on the choice of model physics and stellar modelling code. The MLT is an approximation of convection which is often calibrated to the Sun and then assumed for all stars in a model. However, studies of 3D hydrodynamical simulations suggest that the degree to which $\mlt$ approximates convection varies across the HR diagram \citep{Magic.Weiss.ea2015} and this is confirmed when modelling stars with asteroseismology \citep{Tayar.Somers.ea2017}.

The PP model (without the Sun) favoured a mean mixing-length parameter of $\mu_\alpha \simeq 1.7$. Whereas, the PPS model yielded a higher value of $\mu_\alpha \simeq 1.9$ by $\sim 2$-$\sigma$. We found this was attributed to the addition of the Sun. The individual solar results for the PPS model yielded a value of $\mlt_\odot = 2.12\pm0.03$ which was considerably higher than the $\mlt$ obtained for the other stars in the sample (see Appendix \ref{sec:sun-res}). This result agrees with the 3D simulations \citep[e.g.][]{Trampedach.Stein.ea2014}, which predict lower $\mlt$ for stars with lower $\teff$ and higher $\log g$. However, the solar value also exceeds the reference solar calibrated values of $\approx 1.92$ for the same stellar evolution code \citep{Paxton.Bildsten.ea2011}. This is caused by the differences in adopted solar mixture, atmospheric boundary conditions and the treatment of convective mixing between this work and typical reference values.

Despite the difference in $\mu_\alpha$, the resulting spread in mixing-length for the PPS model $\sigma_\alpha \approx 0.13$ was double that of the PP model to cope with the high solar value. This implies that a large population spread in $\mlt$ could explain the difference we see. In other words, if we assume that the best-fitting $\mlt$ is normally distributed in our population, then the Sun lies within 2-$\sigma$ of the mean, among 95 per cent of all stars in the population. 

There are a few prior studies which look at the spread in $\mlt$ for a population of stars, typically by fitting $\mlt$ as a function of $\metallicity$, $\teff$, and $\log g$ \citep[e.g.][]{Bonaca.Tanner.ea2012,Viani.Basu.ea2018}. For example, results from \citet{Viani.Basu.ea2018} for stellar models including diffusion, predict $\mlt$ in the range \numrange{1.5}{2.3} across our sample. This dispersion would be more compatible with the larger spread obtained by our PPS model. However, in future work we should further investigate how $\mlt$ varies with stellar parameters, as our assumption of a normal distribution may not be accurate.

We found a greater difference in $\mlt$ between the models with and without the Sun when we max-pooled $\mlt$. The MP models yielded a global $\mlt$ in-line with $\mu_\alpha$ from the PP model. However, when we added the Sun, the model yielded $\mlt \approx 2.1$ which is in common with the solar results (see Appendix \ref{sec:sun-res}). This had a similar affect as assuming a solar calibrated value, because the model favoured fitting to data with the best observational precision. The change in $\mlt$ between the MP and MPS models resulted in a mean difference of $\sim 20$ per cent between the individual stellar ages. This is an example of how adopting a solar calibrated value can bias stellar ages. We argue that carefully including the Sun as a part of the population with an intrinsic spread is a better way to calibrate the stellar models than assuming as solar $\mlt$ across the sample.

In all observables except for $L$, the Sun is near the centre of our distribution of stars. However, we found no relationship between $L$ and $\mlt$ in both our NP and PP models. A possible explanation for the difference in $\mlt$ with and without the Sun could be some systematic offset in our observational data for the sample. Here, we point to our choice of spectroscopic $\teff$ which typically underestimates $\teff$ compared to photometric scales, as noted in \citetalias{Serenelli.Johnson.ea2017}. We ran a solar model with an additional parameter, $\Delta \teff = T_{\rm eff, obs} - \teff$ which represents a bias in the observed effective temperature. The estimated covariance between $\Delta \teff$ and $\mlt$ was \SI{0.452}{\kelvin}\footnote{\mlt~is dimensionless, hence the units of covariance are \si{\kelvin}.} (with a correlation of \num{0.517}). Therefore, underestimating $\teff$ by about \SI{100}{\kelvin} could underestimate $\mlt$ by about \num{0.1}. If we extend this result to the other stars in the sample, the lower $\mlt$ obtained without including the Sun as a star could be caused by underestimating $\teff$ relative to the Sun. Alternatively, the $\mlt$ of the Sun could have been higher than the rest of the sample to compensate for neglecting additional sources of mixing required to reproduce the higher precision solar observables. A deeper quantification of systematic uncertainties is left to future work.

\subsection{Comparison with APOKASC results}\label{sec:comp}

Before we compare our results to \citetalias{Serenelli.Johnson.ea2017}, we should highlight some key differences between our data and methodology. The results from \citetalias{Serenelli.Johnson.ea2017} were determined using a grid-based-modelling technique, which estimates the likelihood across a dense grid of stellar models. They used results from several pipelines to estimate the systematic uncertainties. For the central values of their results, they used the Bayesian stellar algorithm \citep[BASTA;][]{SilvaAguirre.Davies.ea2015} using a grid computed with \textsc{GARSTEC} \citep{Weiss.Schlattl2008}. Their choice of stellar physics was similar to this work, except for two major differences.

Firstly, the results of \citetalias{Serenelli.Johnson.ea2017} were determined using stellar models calculated without heavy-element diffusion. The inclusion of diffusion when modelling the Sun has been commonplace over the last few decades, with good agreement between models and helioseismic observations \citep{Christensen-Dalsgaard.Proffitt.ea1993, Bahcall.Pinsonneault.ea1995}. More recent work explored the diffusion in cluster stars \citep{Korn.Grundahl.ea2007, Onehag.Gustafsson.ea2014} and another demonstrated the impact of including diffusion on stellar ages \citep{Dotter.Conroy.ea2017}. Our stellar models were computed with heavy-element diffusion. Recently, work by \citet{Nsamba.Campante.ea2018} on a similar sample of stars showed, on average, models without diffusion compared to those including diffusion can lead to underestimated radii and masses, and overestimated ages by 1, 3, and 16 per cent respectively.

Secondly, our choice of \citet{Asplund.Grevesse.ea2009} solar chemical mixture differs from the \citet{Grevesse.Sauval1998} mixtures adopted by \citetalias{Serenelli.Johnson.ea2017}. The former leads to a solar heavy-element to hydrogen ratio of $(Z/X)_\odot = 0.0181$, and the latter, $(Z/X)_\odot = 0.0230$. Typically, \citet{Grevesse.Sauval1998} abundances are favoured in asteroseismic modelling because they are better able to reproduce measurements of helium in the Sun from helioseismology \citep{Serenelli.Basu.ea2009}. An effect of using the \citet{Asplund.Grevesse.ea2009} abundances, is that it favours lower $Z_\mathrm{init}$ for a given $\metallicity_\mathrm{surf}$. As a result, models using \citet{Grevesse.Sauval1998} abundances on average underestimate radii and mass compared to those without by about 1 and 0.5 per cent respectively \citep{Nsamba.Campante.ea2018}.

Although updated, much of our observable data is comparable to that of \citetalias{Serenelli.Johnson.ea2017}, with the exception of $\teff$. The preferred results from \citetalias{Serenelli.Johnson.ea2017} were determined using a photometric $\teff$ scale which we found to be on average $\sim \SI{170}{\kelvin}$ greater than our spectroscopic scale from DR14. In \citetalias{Serenelli.Johnson.ea2017}, they saw a similar offset between the DR13 $\teff$ available at the time. They found a median difference in mass, radius, and age of approximately $-6$, $-2$, and $+35$ per cent respectively with results from the photometric $\teff$ scale subtracted from the spectroscopic scale.

\begin{figure*}
    \centering
    \begin{subfigure}[b]{.33\linewidth}
        \centering
        \includegraphics[width=\linewidth]{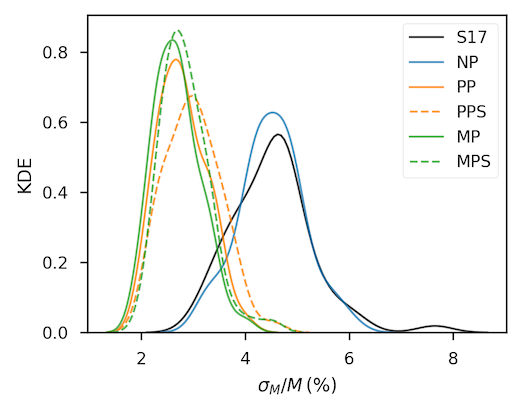}
    \end{subfigure}%
    \begin{subfigure}[b]{.33\linewidth}
        \centering
        \includegraphics[width=\linewidth]{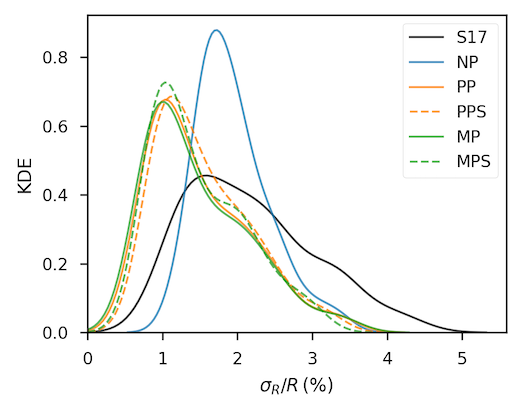}
    \end{subfigure}%
    \begin{subfigure}[b]{.33\linewidth}
        \centering
        \includegraphics[width=\linewidth]{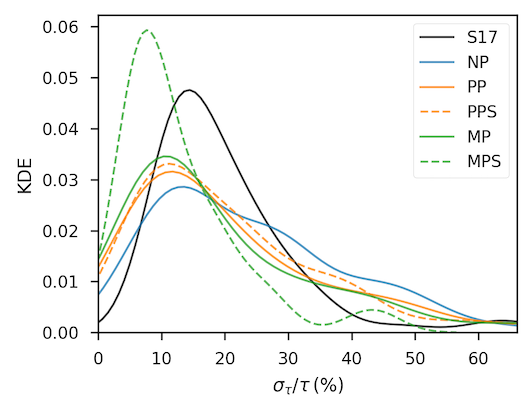}
    \end{subfigure}%
    \caption{Kernel density estimates (KDEs) of the distribution of statistical uncertainties in the results from each model compared with those of \citepalias{Serenelli.Johnson.ea2017} for the sample of APOKASC dwarfs and subgiants.}
    \label{fig:unc-comp}
\end{figure*}

In Fig. \ref{fig:unc-comp} we compare our statistical uncertainties for $M$, $R$, and $\tau$ with those for the equivalent stars from \citetalias{Serenelli.Johnson.ea2017}. We found that the NP model yielded comparable uncertainties to \citetalias{Serenelli.Johnson.ea2017} but note that these are likely underestimated due the influence of the prior boundaries for $Y_\mathrm{init}$ and $\mlt$. We expected larger uncertainties because we included additional free parameters ($Y_\mathrm{init}$ and $\mlt$) over the work of \citetalias{Serenelli.Johnson.ea2017}. However, when we treat these parameters hierarchically, we saw a reduction in uncertainties from all of the pooled models. This is because our prior assumptions about the population allows for the sharing of information between the stars. This uncertainty reduction scales with the number of stars in our sample, demonstrated by the results for the synthetic stars in Fig. \ref{fig:shrinkage}. Thus, hierarchically modelling our population resulted in improved statistical uncertainties in stellar fundamental parameters.

\begin{figure*}
    \centering
    \begin{subfigure}[b]{.33\linewidth}
        \includegraphics[width=\linewidth]{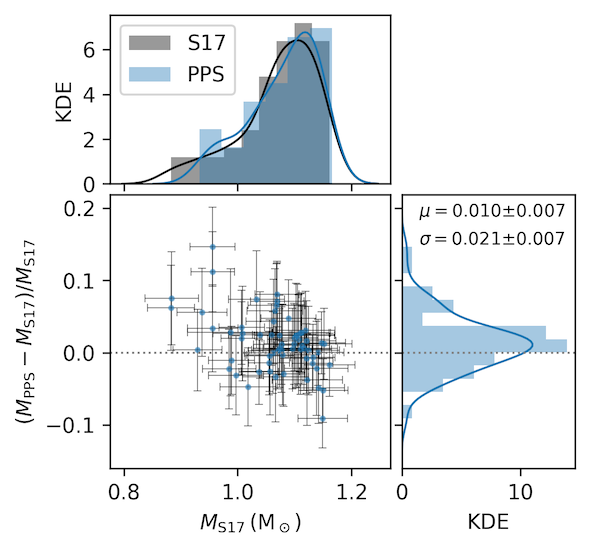}
    \end{subfigure}%
    \begin{subfigure}[b]{.33\linewidth}
        \includegraphics[width=\linewidth]{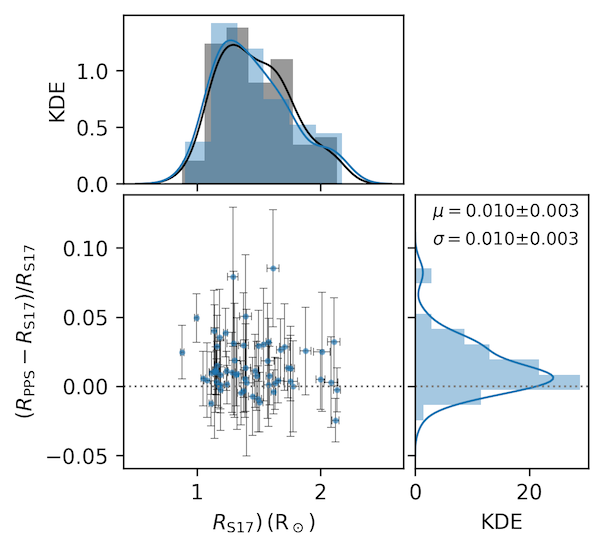}
    \end{subfigure}%
    \begin{subfigure}[b]{.33\linewidth}
        \includegraphics[width=\linewidth]{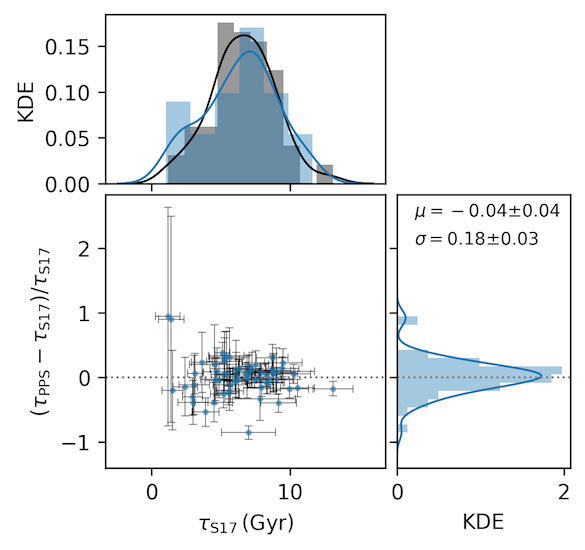}
    \end{subfigure}%
    \caption{The mean and standard deviation in age, mass and radius results from the PPS model compared with the results (using the photometric temperature scale) from \citetalias{Serenelli.Johnson.ea2017}.}
    \label{fig:comp}
\end{figure*}

In the following subsection, we compare the results between our PPS model with that of \citetalias{Serenelli.Johnson.ea2017} for mass, radius, and age with reference to Fig. \ref{fig:comp}. We preferred the PPS model for comparison because it utilised the high-precision data available for the Sun as a star to help calibrate the sample, while partially pooling both $Y_\mathrm{init}$ and $\mlt$ to allow for small variations within the population.

\subsubsection{Mass, radius, and age}

In the left panel of Fig. \ref{fig:comp}, we compare the masses obtained by the PPS model with \citetalias{Serenelli.Johnson.ea2017} and found a dispersion of around 2 per cent. Our masses were on average 1 per cent above the results from \citetalias{Serenelli.Johnson.ea2017}. Although we might expect the lower $\teff$ scale in this work to underestimate the mass, we attribute this overall effect to our choice of stellar model physics. As previously discussed, the use of \citet{Asplund.Grevesse.ea2009} solar abundances and heavy-element diffusion has the cumulative effect of overestimating stellar masses compared to the physics adopted by \citetalias{Serenelli.Johnson.ea2017}. We also found that the results from all the pooled models returned similar masses, with or without the Sun.

In the central panel of Fig. \ref{fig:comp}, we show that our radii are similar to \citetalias{Serenelli.Johnson.ea2017} with a spread of 1 per cent. We also found radii on average 1 per cent greater than the APOKASC results. Similarly to with mass, this contradicts what would be expected from a lower $\teff$ scale and could also be explained by model physics. Our radii also varied little between models with and without the Sun.

Our ages were also consistent with those from \citetalias{Serenelli.Johnson.ea2017}. The right-most panel of Fig. \ref{fig:comp} shows the spread in the relative age differences to be about 18 per cent, slightly underestimated by 4 per cent. We would expect the lower $\teff$ scale to overestimate the ages as found in \citetalias{Serenelli.Johnson.ea2017}, but instead they are comparable. However, as discussed previously, including diffusion has been shown to reduce age estimates compared to those without. Since we have included diffusion, this could explain the similar ages despite the difference in $\teff$ scale.

Including the Sun in our pooled models affected the resulting ages more than mass and radius. Including the Sun typically overestimated the ages compared to models without the Sun. This is expected given the higher $\mlt$ for the models including solar data, because a larger mixing-length leads to more efficient nuclear burning and more time spent during the MS phase. 

\subsection{Systematic uncertainties}\label{sec:sys}

We have already accounted for systematics due to the choice of helium enrichment and mixing-length parameter by marginalising over their uncertainties assuming their population distributions. However, there are other model physics which we have not freely varied, including diffusion and choice of solar mixture. Although our method can be adapted to different stellar evolutionary codes and choice of physics, an in-depth analysis of systematic uncertainties is left to future work. 

In previous work studying stars in the APOKASC sample, several pipelines used a range of stellar evolutionary codes and model physics are employed to evaluate systematic uncertainties from the models \citep{Serenelli.Johnson.ea2017, SilvaAguirre.Lund.ea2017}. Using a hierarchical model in this work enabled us to reduce median statistic uncertainties to 2.5 per cent in mass, 1.2 per cent in radius, and 12 per cent in age. The systematic uncertainty analysis of \citetalias{Serenelli.Johnson.ea2017} found median systematic uncertainties of 3, 1, and 13 per cent in mass, radius, and age respectively. Reducing statistical uncertainties highlights the importance of understanding systematics uncertainties.

Other systematics could arise from observational data. For example, we chose the ASPCAP DR14 $\teff$ scale which was systematically lower than the photometric scale of choice in \citetalias{Serenelli.Johnson.ea2017}. However, our method was still able to recover similar masses, radii, and ages. This could be explained by our choice of stellar model physics, as discussed previously.

\subsection{Outliers}\label{sec:out}

We identified KIC 9025370 as a possible outlier. Consistent across all our models, its output effective temperature, $\teff=\SI{5934(50)}{\kelvin}$ was about 4-$\sigma$ greater than its observed $\teff$, and its modelled $L$ was about 2-$\sigma$ dimmer than its observed luminosity. Only $\dnu$ and $\metallicity_\mathrm{surf}$ were consistent between modelled and observed values. The difference was also apparent in our comparison of ages with \citetalias{Serenelli.Johnson.ea2017} where we obtained an age of $1.5\substack{+0.7\\-0.6}\,\si{\giga\year}$ compared to their value of $7.0\substack{+2.0\\-1.6}\,\si{\giga\year}$.

KIC 9025370 turned out to be a double-lined spectroscopic binary \citep{Nissen.SilvaAguirre.ea2017}, discovered after \citetalias{Serenelli.Johnson.ea2017} and hence included in the original sample. The higher observed luminosity from \textsc{Isoclassify} and inconsistent spectroscopic $\teff$ compared with our model posteriors were compatible with a spectroscopic binary. We calculated a photometric $\teff$ using the IRFM method \citep{Casagrande.Ramirez.ea2010} with the available 2MASS photometry for the target and obtained $\teff=\SI{5983(120)}{\kelvin}$, more consistent with our modelled effective temperature and inconsistent with its spectroscopic $\teff$. Thus, our inferred $\teff$ was within the dispersion between different observed $\teff$ scales. Running the model without KIC 9025370 did not affect the resulting inferred hyperparameters, demonstrating the robustness of our model. Therefore, we present KIC 9025370 in our results but suggest that further investigation should be carried out.

\section{Conclusion}



We have shown that modelling $Y_\mathrm{init}$ and $\mlt$ to improve inference of fundamental parameters can be done through the use of an HBM, whilst still improving statistical uncertainties. Our results were in good agreement with \citetalias{Serenelli.Johnson.ea2017} with small changes in mass and radii expected from our choice of model physics and updated observables. Taking our partially-pooled model including the Sun (PPS) as our preferred set of results, we obtained median statistical uncertainties on $M$, $R$, and $\tau$ of 2.5, 1.2, and 12 per cent respectively. Furthermore, we demonstrated that the uncertainties reduced with increasing sample size in a population of synthetic stars, giving scope to further improve our inference on larger sample sizes from \emph{TESS}.

We found that the gradient, $\Delta Y / \Delta Z$ of the linear helium enrichment law ranged from 0.8 to 1.6 depending on the level of parameter pooling and the inclusion of the Sun in our sample, with $\Delta Y / \Delta Z = 1.1\substack{+0.3\\-0.3}$ from our preferred PPS model. Consistent across our models was the spread in initial helium about the enrichment law, $\sigma_Y = 0.005\substack{+0.004\\-0.003}$. The mean $\mlt$ in the population was $\mu_\alpha = 1.90\substack{+0.10\\-0.09}$ for the PPS model, with values from 1.7 to 2.1 depending on the level of pooling and whether or not solar data was included. We also found the spread in $\mlt$ doubled to $\sigma_\alpha = 0.13\substack{+0.06\\-0.05}$ to account for the addition of the Sun in our sample. We conclude that there are still discrepancies between the best-fitting $\mlt$ in our population and that of the Sun which need to be investigated further. Perhaps, the addition of asteroseismic signatures of helium abundance \citep[see e.g.][]{Verma.Raodeo.ea2017} would improve our constraints on $Y_\mathrm{init}$ and thus reduce star-by-star uncertainties in $\mlt$.

Using HBMs has allowed us to introduce more free parameters without sacrificing statistical uncertainties. We used an ANN to approximate stellar models, a method which can be extended to higher input dimensions with little impact on training and evaluation time. Our model also scales well with the number of stars, making use of GPU parallel processing when sampling the posterior.

As shown in tests with synthetic stars (Appendix \ref{sec:test-stars}) and apparent in Fig. \ref{fig:unc-comp}, increasing the number of stars decreases the statistical uncertainties when parameters are pooled. The theoretical limit to this improvement is $\sqrt{N_1 / N_2}$ for two populations of size $N_1$ and $N_2$. For example, if we increase our sample to 300 stars, we would expect the uncertainties to reduce by up to a factor of 2. Naturally, the uncertainty is still limited by observational precision. However, hierarchical modelling as demonstrated in this work, allows us to get the most out of our data and paves the way for a data-driven analysis of model systematics.

Including all-sky data from \emph{TESS} and in anticipation of \emph{PLATO} \citep{Rauer.Catala.ea2014} we can expect our sample size of asteroseismic dwarfs and subgiants only to increase. There is also scope to extend our grid of models to include red giants, for which there are vast catalogues of stars already studied with \emph{Kepler} \citep{Pinsonneault.Elsworth.ea2018}.

\section*{Acknowledgements}



This work is a part of a project that has received funding from the European Research Council (ERC) under the European Union’s Horizon 2020 research and innovation programme (CartographY; grant agreement ID 804752). A.J.L., G.R.D., and W.J.C. acknowledge the support of the Science and Technology Facilities Council. D.H. acknowledges support from the Alfred P. Sloan Foundation, the National Aeronautics and Space Administration (80NSSC19K0597), and the National Science Foundation (AST-1717000). M.B.N. acknowledges support from the UK Space Agency. R.A.G. acknowledges the funding from the PLATO CNES grant. We acknowledge the use of the \textsc{daft} and \textsc{corner} packages to produce figures in this work. We thank Ilya Mandel and Sean Matt for their discussion regarding this work. We also thank the referee for their immensely helpful comments which have improved the paper.

\section*{Data Availability}



The data underlying this article are available in the article, in its online supplementary material, and in Zenodo, at \url{https://dx.doi.org/10.5281/zenodo.4746353}.




\bibliographystyle{mnras}
\bibliography{references} 



\appendix

\section{Artificial neural network}\label{sec:ann}



Once we constructed our grid of models, we needed a way in which we could continuously sample the grid for use in our statistical model. We opted to train an ANN. The ANN is advantageous over interpolation because it scales well with dimensionality, training and evaluation is fast, and gradient evaluation is easy due to its roots in linear algebra \citep{Haykin2007}. We trained an ANN on the data generated by the grid of stellar models to map fundamentals to observables. Firstly, we split the grid into a \emph{train} and \emph{validation} dataset for tuning the ANN, as described in Appendix \ref{sec:train}. We then tested a multitude of ANN configurations and training data inputs, repeatedly evaluating them with the validation dataset in Appendix \ref{sec:opt}. In Appendix \ref{sec:test}, we reserved a set of randomly generated, off-grid stellar models as our final \emph{test} dataset to evaluate the approximation ability of the best-performing ANN independently from our train and validation data. Here, we briefly describe the theory and motivation behind the ANN.

An ANN is a network of artificial \emph{neurons} which each transform some input vector, $\boldsymbol{x}$ based on trainable weights, $\boldsymbol{w}$ and a bias, $b$. The weights are represented by the connections between neurons and the bias is a unique scalar associated with each neuron. A multi-layered ANN is where neurons are arranged into a series of layers such that any neuron in layer $j-1$ is connected to at least one of the neurons in layer $j$. 

\begin{figure}
    \includegraphics[width=\linewidth]{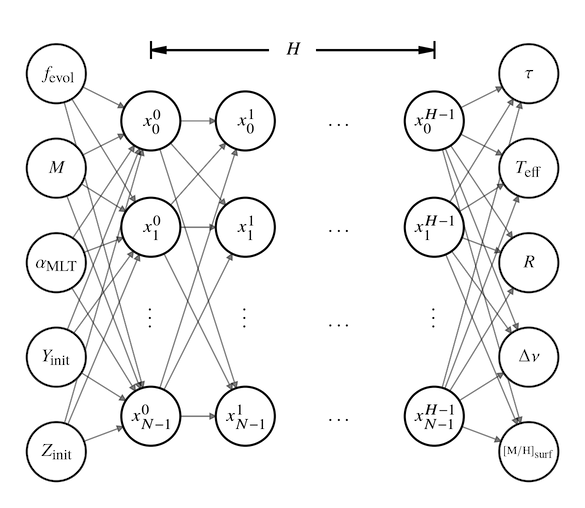}
    \caption{An artificial neural network comprising $H$ hidden layers with $N$ neurons per layer. Arrows connecting the nodes represent tunable weights.}
    \label{fig:net}
\end{figure}

In this work, we considered a fully-connected ANN, where each neuron in layer $j-1$ is connected to every neuron in layer $j$. The output of the $k$-th neuron in layer $j$ is, 
\begin{equation}
    x_{j, k}=f_j(\boldsymbol{w}_{j, k} \cdot \boldsymbol{x}_{j-1} + b_{j, k}),
\end{equation}
where $f_j$ is the \emph{activation} function for the $j$-th layer, $\boldsymbol{w}_{j, k}$ are the weights connecting all the neurons in layer $j-1$ to the current neuron, and $b_{j, k}$ is the bias. This generalises such that the output of the $j$-th layer is,
\begin{equation}
    \boldsymbol{x}_{j}=f_j(\boldsymbol{W}_{j} \cdot \boldsymbol{x}_{j-1} + \boldsymbol{b}_{j}),
\end{equation}
where $\boldsymbol{W}_j$ is the matrix of weights leading to all neurons in the $j$-th layer. For a regression ANN, we typically have a linear activation function applied to the final output layer. Layers of neurons between the input and output layers are called \emph{hidden} layers. Therefore, the output of a network of $H$ hidden layers with initial input $\boldsymbol{\mathbb{X}}$ is,
\begin{equation}
    \widetilde{\boldsymbol{\mathbb{Y}}} = \boldsymbol{W}_{H} \cdot f_{H-1}(\dots f_1(\boldsymbol{W}_1 \cdot f_0(\boldsymbol{W}_{0} \cdot \boldsymbol{\mathbb{X}} + \boldsymbol{b}_{0}) + \boldsymbol{b}_1) ) + \boldsymbol{b}_{H}.
\end{equation}
We also restricted our configuration to an ANN with the same number of neurons, $N$ in each hidden layer. Hereafter, we refer to our choice of neurons per layer, $N$ and hidden layers, $H$ as the \emph{architecture} (see Fig. \ref{fig:net}).

To fit the ANN, we used a set of training data, $\boldsymbol{\mathbb{D}}_\mathrm{train} = \{\boldsymbol{\mathbb{X}}_i, \boldsymbol{\mathbb{Y}}_i\}_{i=1}^{N_\mathrm{train}}$ comprising $N_\mathrm{train}$ input-output pairs. We split the training data into pseudo-random batches, $\boldsymbol{\mathbb{D}}_\mathrm{batch}$ because this has been shown to improve ANN stability and computational efficiency \citep{Masters.Luschi2018}. The set of predictions made for each batch is evaluated using a \emph{loss} function which primarily comprises an error function, $E(\boldsymbol{\mathbb{D}}_\mathrm{batch})$ to quantify the difference between the training data outputs ($\boldsymbol{\mathbb{Y}}$) and predictions ($\widetilde{\boldsymbol{\mathbb{Y}}}$). We also considered an additional term to the loss called \emph{regularisation} which helps reduce over-fitting \citep{Goodfellow.Bengio.ea2016}. During fitting, the weights are updated after each batch using an algorithm called the \emph{optimizer}, back-propagating the error with the goal of minimising the loss such that $\widetilde{\boldsymbol{\mathbb{Y}}} \approx \boldsymbol{\mathbb{Y}}$ \citep[see e.g.][]{Rumelhart.Hinton.ea1986}.

We trained the ANN using \textsc{TensorFlow} \citep{Abadi.Barham.ea2016}. We varied the architecture, number of batches, choice of loss function, optimizer, and regularisation during the optimisation phase. For each set of ANN parameters, we initialised the ANN with a random set of weights and biases and minimized the loss over a given number of \emph{epochs}. An epoch is defined as one iteration through the entire training dataset, $\boldsymbol{\mathbb{D}}_\mathrm{train}$. We tracked the loss for each ANN using an independent validation dataset to determine the most effective choice of ANN parameters (see Appendix \ref{sec:opt}).

\subsection{Train, validation, and test data}\label{sec:train}



We built the train and validation datasets from the outputs of the grid of stellar models in Section \ref{sec:grid}. This included the input parameters: $M$, $\mlt$, $Y_\mathrm{init}$, and the initial heavy-elements fraction, $Z_\mathrm{init}$. We also included the $\teff$, $\log g$, $\dnu$, stellar age ($\tau$), radius ($R$), surface metallicity ($\metallicity_\mathrm{surf}$), and other chemical composition information generated by the models. We determined the fractional MS lifetime, $f_{\mathrm{MS}} = \tau / \tau_{\mathrm{MS}}$, of each evolutionary track by taking $\tau_{\mathrm{MS}}$ as the age when the central hydrogen fraction, $X_c < 0.01$. We then cut data where $f_{\mathrm{MS}} < 0.01$ to remove points on the grid prior to the MS. Once we had refined the data from the grid of models, we randomly sampled \num{7.736e6} points to use as the training dataset, with the remaining $\sim \num{2e6}$ points given to the validation dataset. We varied our choice of ANN input and output parameters among those available in the dataset during tuning (see Appendix \ref{sec:opt}).

Additionally, we produced a test dataset of $\sim \num{2e6}$ stellar models evolved using \textsc{MESA}. Values for the initial $M$, $\metallicity$, $Y$, and $\mlt$ were chosen randomly within the range of grid parameters described in Table \ref{tab:grid} such that they spanned the breadth of the grid in an unbiased manner. We prepared this dataset in the same way as the training set, but also constrained it to $\tau < \SI{15}{\giga\year}$ because we consider ages above $\sim \SI{15}{\giga\year}$ unphysical and such points are sparse in the training data. The test dataset was set aside and evaluated on the final ANN.

\subsection{Tuning}\label{sec:opt}



We trained an ANN to reproduce stellar observables according to our choice of physics with greater accuracy than typical observational precisions. We experimented with a variety of ANN parameter choices, such as the architecture, activation function, optimization algorithm, and loss function. We tuned the ANN parameters by varying them in both a grid-based and heuristic approach, each time evaluating the accuracy using the validation dataset.

During initial tuning, we found that having stellar age as an input was unstable, because it varied heavily with the other input parameters. We mitigated this by introducing an input to describe the fraction of time a star had spent in a given evolutionary phase, $f_\mathrm{evol}$. 
\begin{equation}
    f_\mathrm{evol} = \begin{cases}
        f_\mathrm{MS},\quad &f_\mathrm{MS} \leq 1\\
        1 + \frac{\tau\,-\,\tau_\mathrm{MS}}{\tau_\mathrm{end}\,-\,\tau_\mathrm{MS}},\quad &f_\mathrm{MS} > 1
    \end{cases}\label{eq:fevol}
\end{equation}
where $\tau_\mathrm{end}$ is the age of the star at the end of the track,
\begin{equation}
    f_\mathrm{MS} = \frac{\tau}{\tau_\mathrm{MS}},
\end{equation}
and $\tau_\mathrm{MS}$ is the MS lifetime. A star with $0.01 < f_\mathrm{evol} \leq 1.0$ is in its MS phase, burning hydrogen in its core, and $1.0 < f_\mathrm{evol} \leq 2.0$ has left the MS. Consequently, $f_\mathrm{evol}$ gives the ANN information about the internal state of the star which affects the output observables. Otherwise, $f_\mathrm{evol}$ has little physical meaning, although it could be interpreted as a measure of the evolutionary phase of the star.

We also observed that the ANN trained poorly in areas with a high rate of change in observables, likely because of poor grid coverage in those areas. To bias training to such areas, we calculated the gradient in $\teff$ and $\log g$ between each point for each stellar evolutionary track and used them as optional weights to the loss during tuning. These weights multiplied the difference between the ANN prediction and the training data in our chosen loss function.

After preliminary tuning, we chose the ANN input and output parameters to be $\boldsymbol{\mathbb{X}} = \{f_\mathrm{evol}, M, \mlt, Y_\mathrm{init}, Z_\mathrm{init}\}$ and $\boldsymbol{\mathbb{Y}} = \{\log(\tau), \teff, R, \dnu, \metallicity_\mathrm{surf}\}$ respectively. A generalised form of our neural network is depicted in Fig. \ref{fig:net}. The inputs corresponded to initial conditions in the stellar modelling code and the outputs corresponded to surface conditions throughout the lifetime of the star, with the exception of age which is mapped from $f_\mathrm{evol}$.

We standardised the training dataset by subtracting the median, $\mu_{1/2}$ and dividing by the standard deviation, $\sigma$ for each input and output parameter. We found that the ANN performed better when the training data was scaled in this way as opposed to no scaling at all. We present the parameters used to standardise the training dataset in Table \ref{tab:std}.

\begin{table*}
    \caption{The median, $\mu_{1/2}$ and standard deviation, $\sigma$ for each parameter in the training data, used to standardise the dataset.}
    \label{tab:std}
    \input{tables/standardisation.tex}
\end{table*}

We found that the optimal choice of architecture ($N$ and $H$) varied depending on our choice of other ANN parameters. Therefore, each time we explored a new parameter, we trained an ANN with a grid of $(N,H)$ ranging from $(32, 2)$ to $(512, 10)$.

We evaluated the performance of three activation functions: the hyperbolic-tangent, the rectified linear unit \citep[ReLU;][]{Hahnloser.Sarpeshkar.ea2000, Glorot.Bordes.ea2011} and the exponential linear unit \citep[ELU;][]{Clevert.Unterthiner.ea2015}. Although the ReLU activation function out-performed the other two in speed and accuracy, the resulting ANN output was not smooth. The discontinuity in the ReLU function, $f(x) = \max(0, x)$ in turn caused the output to be discontinuous. This made the ANN difficult to sample for our choice of statistical model (see Section \ref{sec:hbm}). Out of the remaining activation functions, ELU performed the best, providing a smooth output which was well-suited to our probabilistic sampling methods.

We compared the performance of two optimisers: Adam \citep{Kingma.Ba2014} and stochastic gradient descent \citep[SGD; see e.g.][]{Ruder2016} with and without momentum \citep{Qian1999}. Both optimizers required a choice of \emph{learning rate} which determined the rate at which the weights were adjusted during training. We found that Adam performed well but the validation loss was noisy as a function of epochs as it struggled to converge. The SGD optimizer was less noisy than Adam, but it was difficult to tune the learning rate. However, SGD with momentum allowed for more adaptive weight updates and out-performed the other configurations.

There are several ways to reduce over-fitting, from minimising the complexity of the architecture, to increasing the size and coverage of the training dataset. One alternative is to introduce weight regularisation. So-called L2 regularisation adds a term, $\sim \lambda_k \sum_i w_{i, k}^2$ to the loss function for a given hidden layer, $k$ which acts to keep the weights small. We varied the magnitude of $\lambda_k$ and found that if it was too large it would dominate the loss function, but if it was too small then performance on the validation dataset was poorer.

We compared the choice of two error functions: mean squared error (MSE) and mean absolute error (MAE). The former is widely used among ANNs because it is more sensitive to large errors. However, we tracked both metrics regardless of which was added to the loss function and found that MAE converged faster. Although MAE is less effective at large errors, we found that these were typically at the edges of the grid and the accuracy was good enough everywhere else.

\begin{figure}
    \centering
    \includegraphics[width=\linewidth]{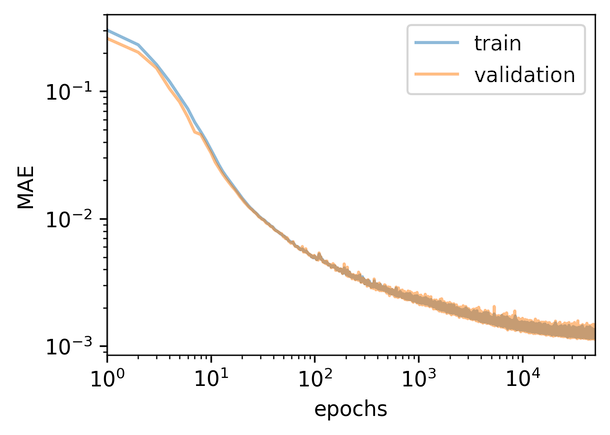}
    \caption{The MAE as a function of epochs for the train and validation datasets.}
    \label{fig:loss}
\end{figure}

After extensive tuning, we opted for an ANN with $N=128$ neurons in each of $H=6$ hidden layers. Each of the hidden layers used an ELU activation function and L2 weight regularisation with $\lambda = \num{1e-6}$. We trained the ANN for \num{50000} epochs with a \num{500} training data batches each containing \num{15472} input-output pairs. To fit the ANN, we used an SGD optimiser with an initial learning rate of \num{1e-4} and momentum of \num{0.999} with an MAE loss function. Training took $\sim \SI{48}{\hour}$ on an NVidia Tesla V100 graphics processing unit (GPU). In Fig. \ref{fig:loss} we show the training and validation MAE as a function of epochs for the final ANN configuration. The training and validation loss were comparable throughout training.

\subsection{Testing}\label{sec:test}



The test dataset contained $\sim \num{2e6}$ stellar models evolved in the same way as the training dataset, but with initial conditions chosen randomly across the range of the grid. We made predictions for the test dataset, deriving luminosity from the output radius and effective temperature, using the final trained ANN as described in Appendix \ref{sec:opt}. We then evaluated the accuracy of the ANN by taking the difference between the test truth and ANN prediction, $x_\mathrm{true} - x_\mathrm{pred}$. 

\begin{table*}
	\centering
	\caption{The median error, $\mu_{1/2}$ and median absolute deviation of the error, $\sigma_\mathrm{MAD} = 1.4826\cdot\mathrm{median}(|E(\mathbb{Y}) - \mu_{1/2}|)$ for a given ANN output parameter, $\mathbb{Y}$ from the test dataset. The error, $E(\mathbb{Y})$, is given in the table header, where $\delta \mathbb{Y} = \widetilde{\mathbb{Y}} - \mathbb{Y}$.}
	\label{tab:test}
    \input{tables/test_random.tex}    
\end{table*}

\begin{figure*}
    \centering
    \includegraphics[width=\linewidth]{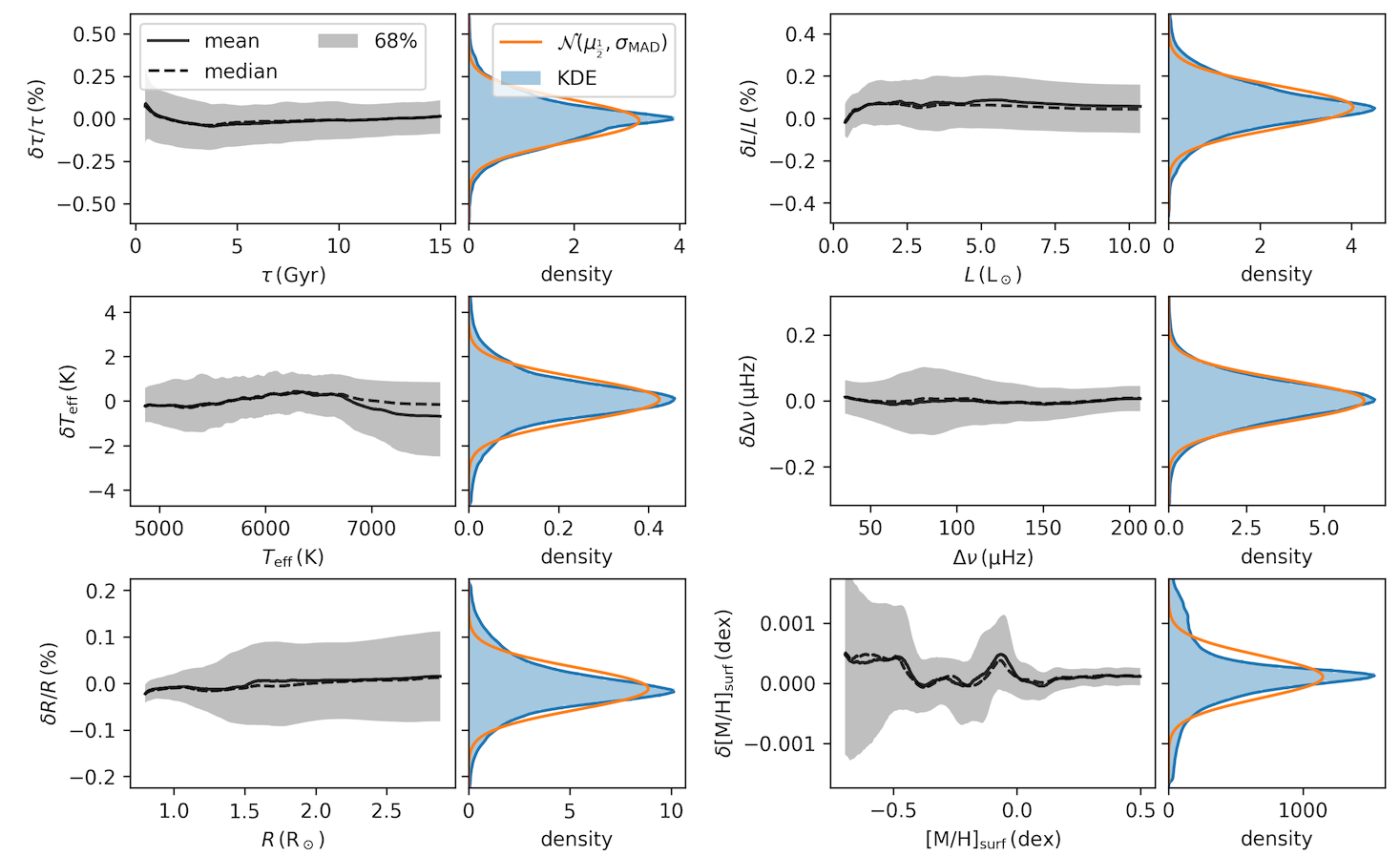}
    \caption{\emph{Left}: the rolling error between a given parameter in the test dataset ($\mathbb{Y}$) and the ANN prediction for that parameter ($\widetilde{\mathbb{Y}}$) where $\delta \mathbb{Y} = \widetilde{\mathbb{Y}} - \mathbb{Y}$. \emph{Right}: a kernel density estimate (KDE) of the error and a normal distribution centred on the median, $\mu_{1/2}$ with an estimator for the standard deviation from the median absolute deviation, $\sigma_\mathrm{MAD}$.}
    \label{fig:test}
\end{figure*}

We found good agreement between the test dataset and ANN predictions, within typical observational uncertainties. We noted that the largest errors lay at the boundaries of the training data and in areas sparsely populated by the grid. This is apparent in Fig. \ref{fig:test} where we plot the test error against each parameter. For example, the spread in error increases for $\metallicity_\mathrm{surf} < -0.5$ where training data is sparse at the edge of the grid. However, the accuracy is very good within the observed range covered by our sample of 81 dwarfs and subgiants. Hence, we chose the median absolute deviation (MAD) as an estimator of the spread in error because it is less sensitive to large errors at the grid boundary than the standard deviation.

To represent the accuracy of the ANN, we present the median, $\mu_{1/2}$ and MAD estimator, $\sigma_\mathrm{MAD} = 1.4826\cdot\mathrm{median}(|E(x) - \mu_{1/2}|)$ of the error ($E(x)$) in Table \ref{tab:test}. The median is close to zero for all parameters, showing little systematic bias in the ANN. The MAD is also lower than observational uncertainties quoted in Section \ref{sec:data}. The spread in error for $\dnu$ of $\SI{0.06}{\mu\Hz}$ is comparable to a small number of observations with the best signal-to-noise. However, the error in $\dnu$ predictions is also comparable to other systematic uncertainties in $\dnu$ discussed in Section \ref{subsec:seismo_model}. Therefore, a robust model which takes account of systematic uncertainties pertaining to $\dnu$, including those from the ANN, will be explored in future work (Carboneau et al. in preparation).

\section{Prior distributions}

\begin{figure}
    \centering
    \includegraphics[width=\linewidth]{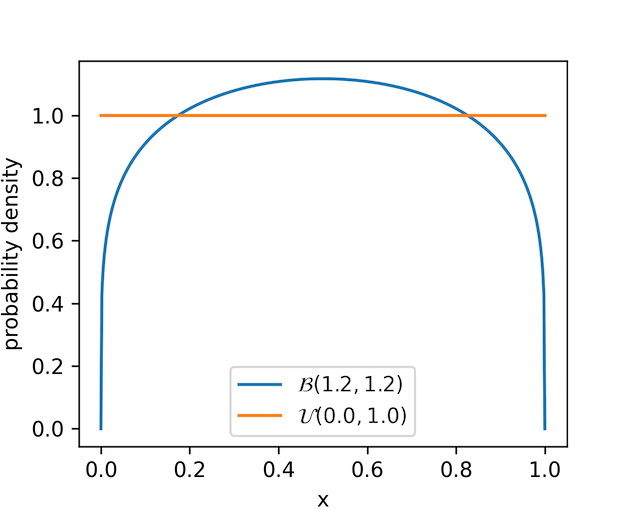}
    \caption{A beta distribution ($\mathcal{B}$) with $\alpha = \beta = 1.2$ for some parameter $x$ and a uniform distribution ($\mathcal{U}$) from 0 to 1.}
    \label{fig:beta}
\end{figure}

We chose a transformed beta distribution (see Equation \ref{eq:beta}) as the prior for the non-pooled stellar parameters as an alternative to a uniform distribution. Fig. \ref{fig:beta} shows the beta distribution compared with a uniform distribution for some parameter $x$ from 0 to 1. We found that the continuously differentiable nature of the beta distribution was preferred by the NUTS over the uniform distribution.

\section{The synthetic population}\label{sec:test-stars}



In this section, we present the results for the NP, PP, and MP models run on a synthetic sample of 100 stars with the following initial conditions. We randomly generated initial $M$ and $\metallicity_\mathrm{init}$ uniformly. We drew initial values for $Y_\mathrm{init}$ from a normal distribution centred on the helium enrichment law from Equation \ref{eq:helium} with $\Delta Y / \Delta Z = 1.8$ and $Y_P = 0.247$, and scaled by $\sigma_Y = 0.008$. We also generated initial values for $\mlt$ from a normal distribution centred on $\mu_\alpha = 2.0$ and scaled by $\sigma_\alpha = 0.08$.

We evolved the synthetic stars to randomly chosen ages using \textsc{MESA}. We then took the output $\tau$, $\teff$, $L$, $\dnu$, and $\metallicity_\mathrm{surf}$ from the models and used these as true values for each of the stars. We added random noise to the observed quantities centred on the true values with a standard deviation of 2.2 per cent in $\teff$, 3.5 per cent in $L$, \SI{0.9}{\micro\hertz} in $\dnu$, and \SI{0.07}{\dex} in $\metallicity_\mathrm{surf}$ chosen to be representative of the APOKASC sample.

\subsection{Stellar parameters}

We found that the NP model recovered the true values for the individual stellar parameters, but the uncertainties were unreliable. The observational quantities alone were not good enough to constrain $Y_\mathrm{init}$ and $\mlt$. As a result, their distributions were truncated at the bounds of their priors. These boundary effects skewed the marginalised posterior means for $Y_\mathrm{init}$ and $\mlt$ towards the centre of the prior (0.28 and 2.0 respectively).

The PP model recovered true values for the synthetic stars with more reliable uncertainty than the NP model. The addition of pooling $Y_\mathrm{init}$ and $\mlt$ between the stars improved their uncertainty which reduced the effects of the prior as seen in the NP model. We found little difference between the results of the PP and MP models.

\begin{figure*}
    \centering
    \includegraphics[width=\textwidth]{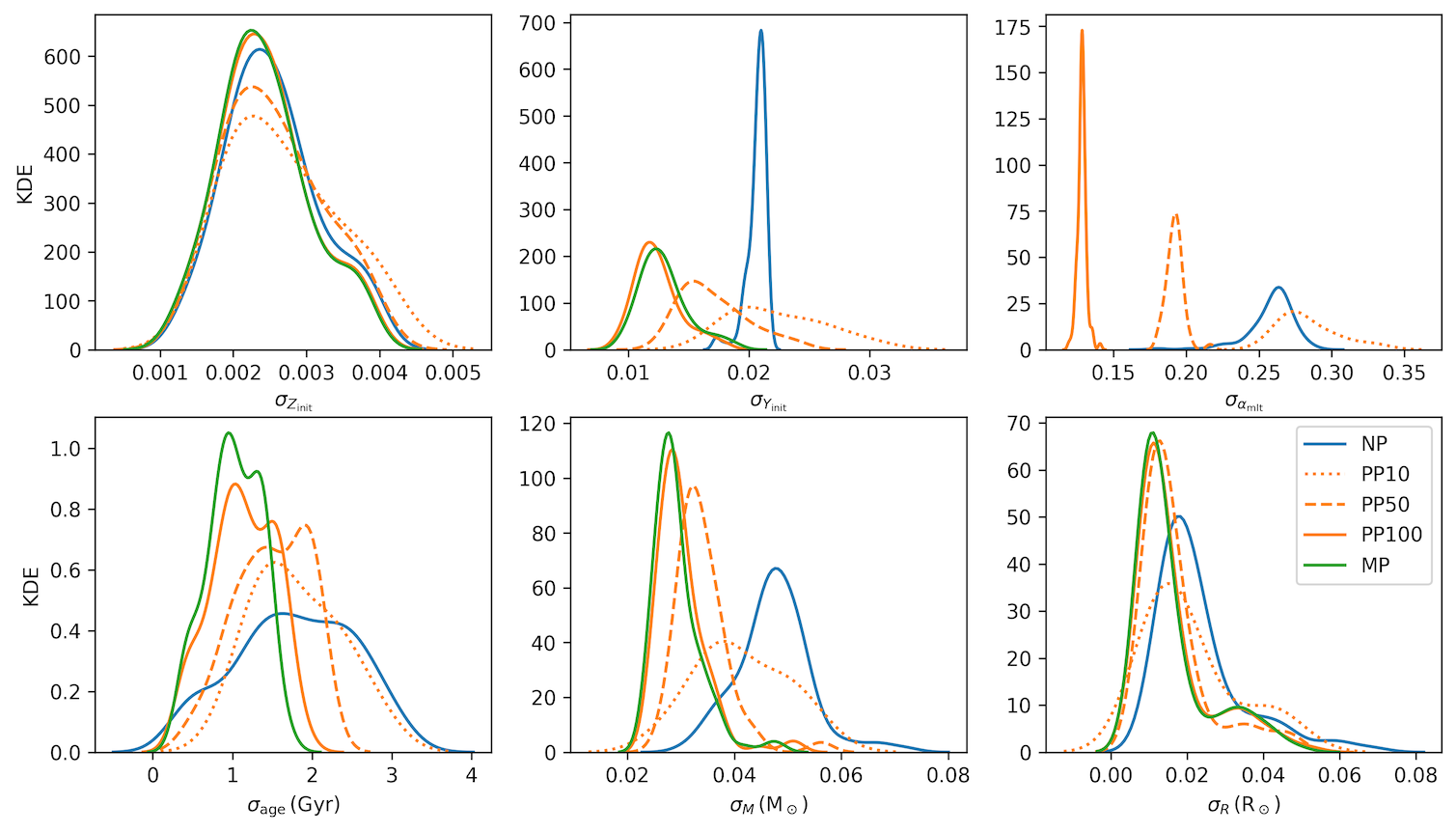}
    \caption{Kernel density estimates (KDEs) of the distributions of statistical uncertainties from each model for the sample of synthetic stars. The PP model was run with 10, 50 and 100 stars and is denoted PP10, PP50, and PP100 respectively. The NP and MP models were both run with the full set of 100 stars.}
    \label{fig:shrinkage}
\end{figure*}

We reran the PP model with 10 and 50 stars chosen randomly from the sample of synthetic stars. In Fig. \ref{fig:shrinkage}, we show the uncertainties in the several parameters from the results of each of the models. For the two pooled parameters, $Y_\mathrm{init}$ and $\mlt$, the uncertainty reduction due to pooling is most obvious. We see the PP model repeatedly improves on the uncertainties from the NP model when $N_\mathrm{stars}$ is increased. 

In Fig. \ref{fig:shrinkage} we also see a similar reduction in uncertainty for $\tau$, $M$, and $R$, with all models improve upon the NP model. However, we do not see the same effect in $Z_\mathrm{init}$ for which the uncertainty appears dominated by observations of $\metallicity_\mathrm{surf}$ .

\subsection{Population parameters}

\begin{figure}
    \centering
    \includegraphics[width=\linewidth]{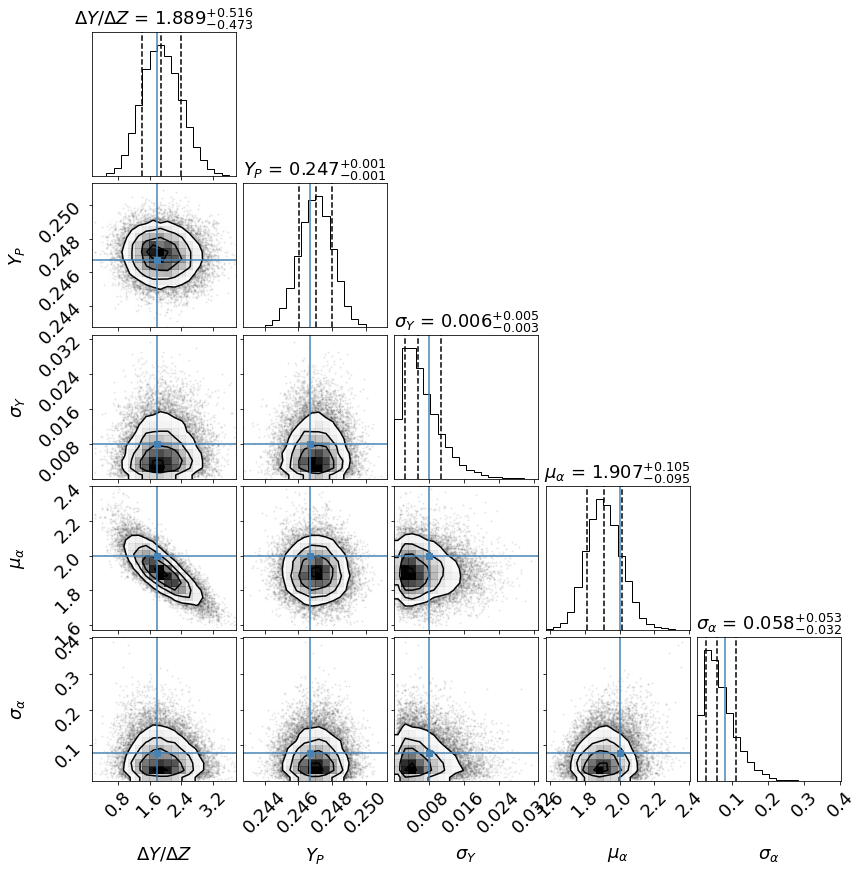}
    \caption{Corner plot showing the marginalised and joint posterior distributions between the PP model hyperparameters for 100 synthetic stars. The true values are shown by the blue lines.}  
    \label{fig:test-corners-pp}   
\end{figure}

In Fig. \ref{fig:test-corners-pp}, we see that the PP model also recovers the hyperparameter truths well, with some noise due to random realisation error. Fitting the model this way has the added benefit over the NP model of improving the inference of the individual stellar parameters, as shown in the previous two sections. We also found that when we ran the PP model with 10 and 50 stars, the uncertainties on the hyperparameters also shrank with increasing $N_\mathrm{stars}$.

\begin{figure}
    \centering
    \includegraphics[width=\linewidth]{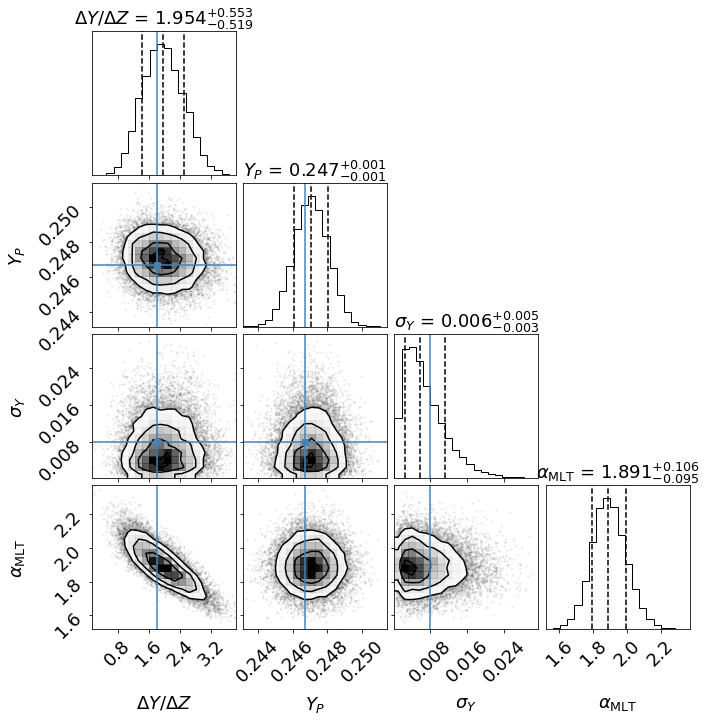}
    \caption{The same as Fig. \ref{fig:test-corners-pp} but for the MP model.}
    \label{fig:test-corners-mp} 
\end{figure}

Fig. \ref{fig:test-corners-mp} shows the hyperparameter results for the MP model. Here, $\mlt$ was assumed to be the same for all stars. This model also recovers the true hyperparameters for helium well, and the assumed value for $\mlt$ is within uncertainty of the true $\mu_\alpha$.

\section{The Sun as a star}\label{sec:sun-res}



\begin{figure*}
    \centering
    \includegraphics[width=\textwidth]{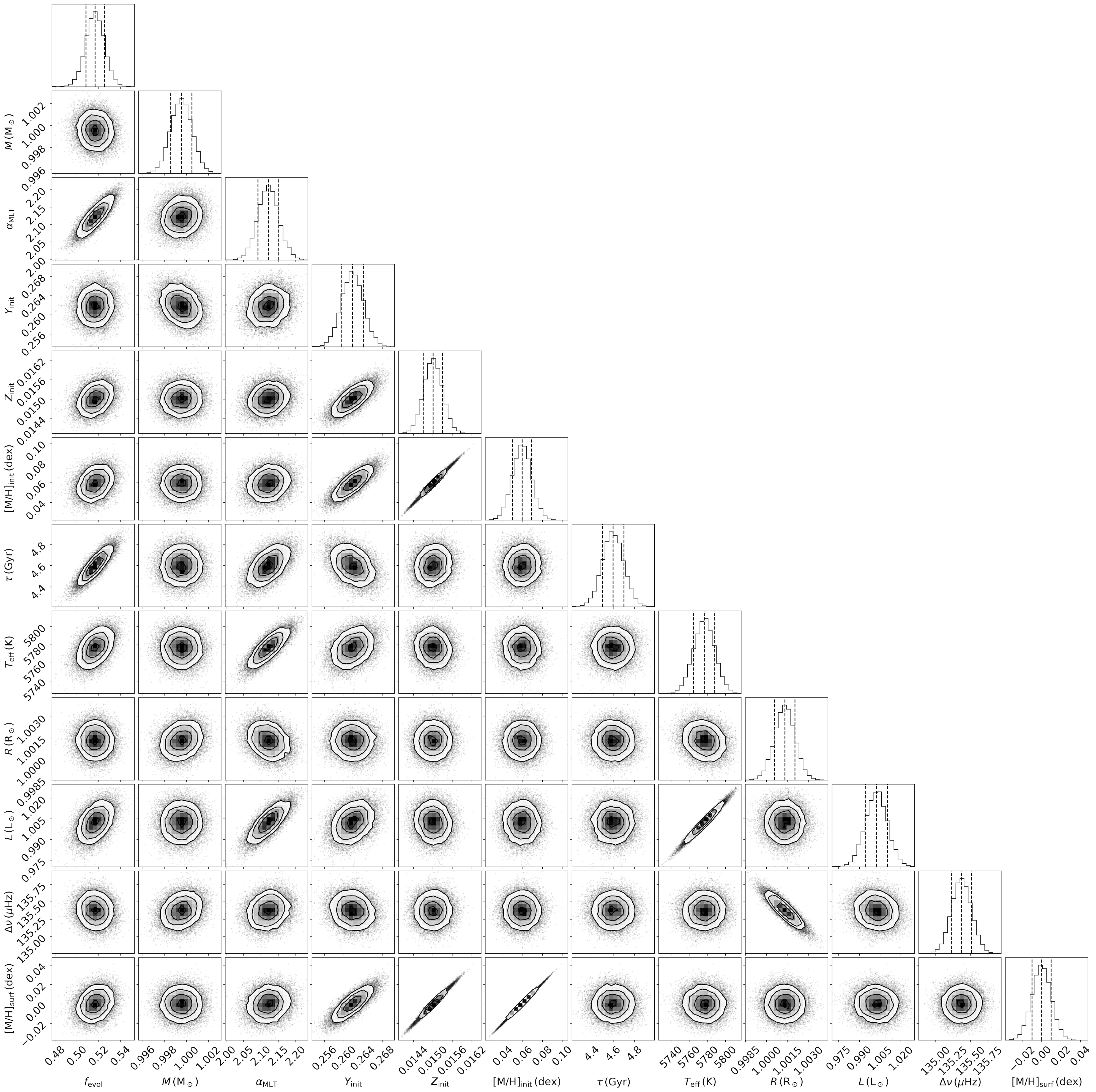}
    \caption{A corner plot showing the sampled marginal and joint posterior distributions for the Sun as a part of the NP model.}
    \label{fig:sun-results}
\end{figure*}

Our model consistently recovers the Sun when modelled in each of the NP, PP, and MP models. In Table \ref{tab:sun-out} we present the results for the Sun as a star from the NP model to show what we obtain without the influence of any other stars in the sample. We show the marginal and joint posterior distributions for the solar parameters from the NP model in the corner plot in Fig. \ref{fig:sun-results}.

We found some differences between $\mlt$ from our solar model and solar calibrations in the literature produced using \textsc{MESA} with similar input physics. For example, the solar calibration in \citet{Stancliffe.Fossati.ea2016} using \citet{Asplund.Grevesse.ea2009} abundances yields compatible initial abundances, $Z_\mathrm{init} = 0.0149$ and $Y_\mathrm{init} = 0.266$ but $\mlt = 1.783$ which differs from our results by about 10-$\sigma$. This is likely because of a few differences in observed values used for the calibration. \citet{Stancliffe.Fossati.ea2016} used observed helium abundance and convection zone depth measurements from helioseismology. Furthermore, solar calibrations typically include convective envelope overshooting. Presuming overshooting increases mixing in the star, we might expect a lower $\mlt$ to compensate this. Therefore, we stress that the addition of the Sun as a star in our model is with the assumption of our choice of input physics.

\begin{table*}
    \centering
    \caption{Solar results from the NP model. The second column shows the median marginalised posterior samples for each parameter with their respective upper and lower 68 per cent credible intervals.}
    \label{tab:sun-out}
    \input{tables/sun_outputs.tex}
\end{table*}

\bsp	
\label{lastpage}
\end{document}

%% file: tables/stars_inputs.tex
\begin{tabular}{lcccc}
\toprule
Name & $\teff\,(\si{\kelvin})$ & $L\,(\si{\solarluminosity})$ & $\dnu\,(\si{\micro\hertz})$ & $\metallicity_\mathrm{surf}\,(\si{\dex})$ \\
\midrule
  KIC10079226 &            $5929\pm125$ &              $1.571\pm0.049$ &             $116.04\pm0.73$ &                           $0.159\pm0.074$ \\
  KIC10215584 &            $5667\pm119$ &              $1.637\pm0.063$ &             $115.16\pm2.83$ &                           $0.043\pm0.069$ \\
  KIC10319352 &            $5456\pm107$ &              $1.848\pm0.056$ &              $78.75\pm1.73$ &                           $0.265\pm0.065$ \\
  KIC10322381 &            $6147\pm149$ &              $2.445\pm0.079$ &              $86.64\pm6.57$ &                          $-0.317\pm0.079$ \\
  KIC10417911 &            $5628\pm110$ &              $3.408\pm0.115$ &              $56.14\pm2.10$ &                           $0.336\pm0.068$ \\
\bottomrule
\end{tabular}

%% file: tables/sun_inputs.tex
\begin{tabular}{lcc}
\toprule
                            Input &            Value &                         Reference \\
\midrule
                    $M\,(\si{\solarmass})$ &  $1.000\pm0.001$ &                               --- \\
                 $\tau\,(\si{\giga\year})$ &      $4.6\pm0.1$ &  \citet{Connelly.Bizzarro.ea2012} \\
                   $\teff\,(\si{\kelvin})$ &      $5777\pm20$ &     \citet{Scott.Grevesse.ea2015} \\
                  $R\,(\si{\solarradius})$ &  $1.000\pm0.001$ &                               --- \\
              $L\,(\si{\solarluminosity})$ &    $1.00\pm0.01$ &                               --- \\
               $\dnu\,(\si{\micro\hertz})$ &    $135.1\pm0.2$ &      \citet{Huber.Bedding.ea2011} \\
 $\metallicity_\mathrm{surf}\,(\si{\dex})$ &    $0.00\pm0.01$ &   \citet{Asplund.Grevesse.ea2009} \\
\bottomrule
\end{tabular}

%% file: tables/stars_outputs_NP.tex
\begin{tabular}{lcccccccccc}
\toprule
Name &                $f_\mathrm{evol}$ &           $M\,(\si{\solarmass})$ &                           $\mlt$ &                $Y_\mathrm{init}$ &                   $Z_\mathrm{init}$ &     $\tau\,(\si{\giga\year})$ &      $\teff\,(\si{\kelvin})$ &         $R\,(\si{\solarradius})$ &     $\dnu\,(\si{\micro\hertz})$ & $\metallicity_\mathrm{surf}\,(\si{\dex})$ \\
\midrule
  KIC10079226 &  $0.44\substack{+0.16 \\ -0.20}$ &  $1.14\substack{+0.04 \\ -0.04}$ &  $2.07\substack{+0.26 \\ -0.30}$ &  $0.28\substack{+0.02 \\ -0.02}$ &  $0.021\substack{+0.003 \\ -0.003}$ &  $2.5\substack{+1.2 \\ -1.3}$ &  $5990\substack{+51 \\ -52}$ &  $1.16\substack{+0.01 \\ -0.02}$ &  $116.0\substack{+0.7 \\ -0.7}$ &           $0.16\substack{+0.07 \\ -0.07}$ \\
  KIC10215584 &  $0.50\substack{+0.21 \\ -0.21}$ &  $1.13\substack{+0.04 \\ -0.05}$ &  $1.92\substack{+0.33 \\ -0.26}$ &  $0.27\substack{+0.02 \\ -0.02}$ &  $0.018\substack{+0.002 \\ -0.002}$ &  $2.9\substack{+1.6 \\ -1.3}$ &  $5949\substack{+64 \\ -65}$ &  $1.18\substack{+0.02 \\ -0.02}$ &  $112.5\substack{+2.6 \\ -2.6}$ &           $0.08\substack{+0.06 \\ -0.06}$ \\
  KIC10319352 &  $1.51\substack{+0.15 \\ -0.28}$ &  $1.09\substack{+0.05 \\ -0.05}$ &  $1.87\substack{+0.33 \\ -0.23}$ &  $0.28\substack{+0.03 \\ -0.02}$ &  $0.028\substack{+0.004 \\ -0.003}$ &  $9.6\substack{+1.7 \\ -1.5}$ &  $5507\substack{+57 \\ -56}$ &  $1.49\substack{+0.03 \\ -0.03}$ &   $78.6\substack{+1.6 \\ -1.6}$ &           $0.28\substack{+0.06 \\ -0.06}$ \\
  KIC10322381 &  $0.98\substack{+0.23 \\ -0.28}$ &  $1.07\substack{+0.07 \\ -0.07}$ &  $2.04\substack{+0.29 \\ -0.31}$ &  $0.28\substack{+0.02 \\ -0.02}$ &  $0.010\substack{+0.002 \\ -0.002}$ &  $4.7\substack{+1.5 \\ -1.7}$ &  $6132\substack{+94 \\ -94}$ &  $1.39\substack{+0.05 \\ -0.04}$ &   $85.6\substack{+4.9 \\ -4.6}$ &          $-0.30\substack{+0.07 \\ -0.08}$ \\
  KIC10732098 &  $1.60\substack{+0.14 \\ -0.19}$ &  $1.12\substack{+0.04 \\ -0.05}$ &  $1.86\substack{+0.33 \\ -0.23}$ &  $0.28\substack{+0.02 \\ -0.02}$ &  $0.017\substack{+0.002 \\ -0.002}$ &  $6.7\substack{+0.8 \\ -0.8}$ &  $5720\substack{+67 \\ -66}$ &  $1.77\substack{+0.04 \\ -0.04}$ &   $62.1\substack{+1.8 \\ -1.7}$ &           $0.06\substack{+0.07 \\ -0.07}$ \\
\bottomrule
\end{tabular}

%% file: tables/stars_outputs_PP.tex
\begin{tabular}{lcccccccccc}
\toprule
Name &                $f_\mathrm{evol}$ &           $M\,(\si{\solarmass})$ &                           $\mlt$ &                $Y_\mathrm{init}$ &                   $Z_\mathrm{init}$ &     $\tau\,(\si{\giga\year})$ &      $\teff\,(\si{\kelvin})$ &         $R\,(\si{\solarradius})$ &     $\dnu\,(\si{\micro\hertz})$ & $\metallicity_\mathrm{surf}\,(\si{\dex})$ \\
\midrule
  KIC10079226 &  $0.22\substack{+0.10 \\ -0.09}$ &  $1.16\substack{+0.02 \\ -0.03}$ &  $1.75\substack{+0.11 \\ -0.09}$ &  $0.28\substack{+0.01 \\ -0.01}$ &  $0.020\substack{+0.003 \\ -0.002}$ &  $1.2\substack{+0.6 \\ -0.5}$ &  $5962\substack{+44 \\ -42}$ &  $1.17\substack{+0.01 \\ -0.01}$ &  $115.9\substack{+0.7 \\ -0.7}$ &           $0.15\substack{+0.07 \\ -0.07}$ \\
  KIC10215584 &  $0.37\substack{+0.15 \\ -0.13}$ &  $1.14\substack{+0.03 \\ -0.03}$ &  $1.74\substack{+0.10 \\ -0.09}$ &  $0.27\substack{+0.01 \\ -0.01}$ &  $0.018\substack{+0.002 \\ -0.002}$ &  $2.1\substack{+1.0 \\ -0.8}$ &  $5941\substack{+57 \\ -56}$ &  $1.18\substack{+0.02 \\ -0.02}$ &  $112.5\substack{+2.6 \\ -2.7}$ &           $0.07\substack{+0.07 \\ -0.07}$ \\
  KIC10319352 &  $1.41\substack{+0.11 \\ -0.27}$ &  $1.08\substack{+0.03 \\ -0.03}$ &  $1.73\substack{+0.10 \\ -0.09}$ &  $0.29\substack{+0.02 \\ -0.01}$ &  $0.028\substack{+0.004 \\ -0.004}$ &  $8.6\substack{+1.1 \\ -1.0}$ &  $5512\substack{+45 \\ -46}$ &  $1.49\substack{+0.02 \\ -0.02}$ &   $78.6\substack{+1.6 \\ -1.6}$ &           $0.28\substack{+0.06 \\ -0.07}$ \\
  KIC10322381 &  $0.78\substack{+0.23 \\ -0.19}$ &  $1.14\substack{+0.03 \\ -0.06}$ &  $1.75\substack{+0.10 \\ -0.09}$ &  $0.26\substack{+0.01 \\ -0.01}$ &  $0.011\substack{+0.002 \\ -0.002}$ &  $3.6\substack{+1.7 \\ -1.1}$ &  $6081\substack{+95 \\ -92}$ &  $1.41\substack{+0.05 \\ -0.05}$ &   $86.2\substack{+4.8 \\ -5.2}$ &          $-0.31\substack{+0.07 \\ -0.07}$ \\
  KIC10732098 &  $1.50\substack{+0.13 \\ -0.14}$ &  $1.14\substack{+0.03 \\ -0.04}$ &  $1.74\substack{+0.10 \\ -0.09}$ &  $0.28\substack{+0.01 \\ -0.01}$ &  $0.018\substack{+0.002 \\ -0.002}$ &  $6.4\substack{+0.6 \\ -0.6}$ &  $5701\substack{+59 \\ -58}$ &  $1.78\substack{+0.03 \\ -0.03}$ &   $62.2\substack{+1.7 \\ -1.7}$ &           $0.06\substack{+0.06 \\ -0.06}$ \\
\bottomrule
\end{tabular}

%% file: tables/stars_outputs_PPS.tex
\begin{tabular}{lccccccccccc}
\toprule
Name &                $f_\mathrm{evol}$ &           $M\,(\si{\solarmass})$ &                           $\mlt$ &                $Y_\mathrm{init}$ &                   $Z_\mathrm{init}$ &     $\tau\,(\si{\giga\year})$ &      $\teff\,(\si{\kelvin})$ &         $R\,(\si{\solarradius})$ &     $\dnu\,(\si{\micro\hertz})$ & $\metallicity_\mathrm{surf}\,(\si{\dex})$ \\
\midrule
  KIC10079226 &  $0.35\substack{+0.11 \\ -0.12}$ &  $1.17\substack{+0.02 \\ -0.03}$ &  $1.95\substack{+0.14 \\ -0.15}$ &  $0.27\substack{+0.01 \\ -0.01}$ &  $0.020\substack{+0.003 \\ -0.002}$ &  $2.1\substack{+0.8 \\ -0.8}$ &  $5962\substack{+44 \\ -43}$ &  $1.17\substack{+0.01 \\ -0.01}$ &  $116.0\substack{+0.7 \\ -0.7}$ &           $0.15\substack{+0.06 \\ -0.07}$ \\
  KIC10215584 &  $0.47\substack{+0.16 \\ -0.16}$ &  $1.14\substack{+0.03 \\ -0.03}$ &  $1.90\substack{+0.15 \\ -0.17}$ &  $0.27\substack{+0.01 \\ -0.01}$ &  $0.018\substack{+0.002 \\ -0.002}$ &  $2.7\substack{+1.2 \\ -1.1}$ &  $5943\substack{+56 \\ -58}$ &  $1.18\substack{+0.02 \\ -0.02}$ &  $112.6\substack{+2.6 \\ -2.6}$ &           $0.07\substack{+0.06 \\ -0.07}$ \\
  KIC10319352 &  $1.51\substack{+0.10 \\ -0.22}$ &  $1.09\substack{+0.03 \\ -0.03}$ &  $1.88\substack{+0.16 \\ -0.16}$ &  $0.28\substack{+0.01 \\ -0.01}$ &  $0.028\substack{+0.004 \\ -0.004}$ &  $9.6\substack{+1.1 \\ -1.2}$ &  $5507\substack{+47 \\ -48}$ &  $1.49\substack{+0.02 \\ -0.02}$ &   $78.6\substack{+1.6 \\ -1.6}$ &           $0.28\substack{+0.06 \\ -0.06}$ \\
  KIC10322381 &  $0.89\substack{+0.21 \\ -0.22}$ &  $1.12\substack{+0.05 \\ -0.06}$ &  $1.93\substack{+0.15 \\ -0.16}$ &  $0.26\substack{+0.01 \\ -0.01}$ &  $0.010\substack{+0.002 \\ -0.002}$ &  $4.3\substack{+1.7 \\ -1.2}$ &  $6093\substack{+92 \\ -89}$ &  $1.41\substack{+0.04 \\ -0.04}$ &   $86.1\substack{+5.0 \\ -4.9}$ &          $-0.31\substack{+0.07 \\ -0.08}$ \\
  KIC10732098 &  $1.60\substack{+0.11 \\ -0.14}$ &  $1.14\substack{+0.03 \\ -0.04}$ &  $1.90\substack{+0.15 \\ -0.17}$ &  $0.27\substack{+0.01 \\ -0.01}$ &  $0.017\substack{+0.002 \\ -0.002}$ &  $6.9\substack{+0.6 \\ -0.6}$ &  $5704\substack{+62 \\ -61}$ &  $1.78\substack{+0.04 \\ -0.03}$ &   $62.2\substack{+1.8 \\ -1.7}$ &           $0.06\substack{+0.06 \\ -0.06}$ \\
\bottomrule
\end{tabular}

%% file: tables/stars_outputs_MP.tex
\begin{tabular}{lccccccccc}
\toprule
Name &                $f_\mathrm{evol}$ &           $M\,(\si{\solarmass})$ &                $Y_\mathrm{init}$ &                   $Z_\mathrm{init}$ &     $\tau\,(\si{\giga\year})$ &      $\teff\,(\si{\kelvin})$ &         $R\,(\si{\solarradius})$ &     $\dnu\,(\si{\micro\hertz})$ & $\metallicity_\mathrm{surf}\,(\si{\dex})$ \\
\midrule
  KIC10079226 &  $0.20\substack{+0.08 \\ -0.08}$ &  $1.17\substack{+0.02 \\ -0.03}$ &  $0.28\substack{+0.01 \\ -0.01}$ &  $0.019\substack{+0.003 \\ -0.002}$ &  $1.1\substack{+0.5 \\ -0.4}$ &  $5961\substack{+42 \\ -41}$ &  $1.17\substack{+0.01 \\ -0.01}$ &  $115.9\substack{+0.7 \\ -0.7}$ &           $0.15\substack{+0.06 \\ -0.07}$ \\
  KIC10215584 &  $0.36\substack{+0.14 \\ -0.13}$ &  $1.14\substack{+0.03 \\ -0.03}$ &  $0.27\substack{+0.01 \\ -0.01}$ &  $0.018\substack{+0.002 \\ -0.002}$ &  $2.0\substack{+0.9 \\ -0.8}$ &  $5941\substack{+57 \\ -57}$ &  $1.18\substack{+0.02 \\ -0.02}$ &  $112.5\substack{+2.6 \\ -2.7}$ &           $0.07\substack{+0.06 \\ -0.07}$ \\
  KIC10319352 &  $1.41\substack{+0.10 \\ -0.25}$ &  $1.08\substack{+0.03 \\ -0.03}$ &  $0.29\substack{+0.02 \\ -0.01}$ &  $0.028\substack{+0.004 \\ -0.004}$ &  $8.6\substack{+1.0 \\ -0.9}$ &  $5512\substack{+44 \\ -45}$ &  $1.49\substack{+0.02 \\ -0.02}$ &   $78.6\substack{+1.7 \\ -1.6}$ &           $0.28\substack{+0.06 \\ -0.07}$ \\
  KIC10322381 &  $0.77\substack{+0.23 \\ -0.19}$ &  $1.14\substack{+0.04 \\ -0.06}$ &  $0.27\substack{+0.01 \\ -0.01}$ &  $0.011\substack{+0.002 \\ -0.002}$ &  $3.5\substack{+1.6 \\ -1.0}$ &  $6076\substack{+96 \\ -91}$ &  $1.41\substack{+0.05 \\ -0.05}$ &   $86.1\substack{+4.7 \\ -5.3}$ &          $-0.32\substack{+0.07 \\ -0.07}$ \\
  KIC10732098 &  $1.50\substack{+0.13 \\ -0.13}$ &  $1.14\substack{+0.03 \\ -0.04}$ &  $0.28\substack{+0.01 \\ -0.01}$ &  $0.018\substack{+0.002 \\ -0.002}$ &  $6.4\substack{+0.6 \\ -0.6}$ &  $5702\substack{+56 \\ -58}$ &  $1.78\substack{+0.03 \\ -0.03}$ &   $62.2\substack{+1.7 \\ -1.7}$ &           $0.06\substack{+0.06 \\ -0.06}$ \\
\bottomrule
\end{tabular}

%% file: tables/stars_outputs_MPS.tex
\begin{tabular}{lccccccccc}
\toprule
Name &                $f_\mathrm{evol}$ &           $M\,(\si{\solarmass})$ &                $Y_\mathrm{init}$ &                   $Z_\mathrm{init}$ &      $\tau\,(\si{\giga\year})$ &      $\teff\,(\si{\kelvin})$ &         $R\,(\si{\solarradius})$ &     $\dnu\,(\si{\micro\hertz})$ & $\metallicity_\mathrm{surf}\,(\si{\dex})$ \\
\midrule
  KIC10079226 &  $0.44\substack{+0.07 \\ -0.06}$ &  $1.16\substack{+0.02 \\ -0.03}$ &  $0.26\substack{+0.01 \\ -0.01}$ &  $0.021\substack{+0.003 \\ -0.002}$ &   $2.7\substack{+0.5 \\ -0.4}$ &  $5965\substack{+40 \\ -40}$ &  $1.17\substack{+0.01 \\ -0.01}$ &  $116.0\substack{+0.7 \\ -0.7}$ &           $0.15\substack{+0.06 \\ -0.06}$ \\
  KIC10215584 &  $0.59\substack{+0.11 \\ -0.13}$ &  $1.13\substack{+0.03 \\ -0.03}$ &  $0.26\substack{+0.01 \\ -0.01}$ &  $0.018\substack{+0.002 \\ -0.002}$ &   $3.6\substack{+0.9 \\ -0.9}$ &  $5952\substack{+55 \\ -56}$ &  $1.18\substack{+0.02 \\ -0.02}$ &  $112.7\substack{+2.7 \\ -2.7}$ &           $0.08\substack{+0.06 \\ -0.07}$ \\
  KIC10319352 &  $1.61\substack{+0.04 \\ -0.06}$ &  $1.08\substack{+0.03 \\ -0.03}$ &  $0.27\substack{+0.01 \\ -0.01}$ &  $0.028\substack{+0.004 \\ -0.003}$ &  $10.8\substack{+0.7 \\ -0.8}$ &  $5516\substack{+46 \\ -47}$ &  $1.49\substack{+0.02 \\ -0.02}$ &   $78.6\substack{+1.7 \\ -1.6}$ &           $0.28\substack{+0.07 \\ -0.06}$ \\
  KIC10322381 &  $0.98\substack{+0.19 \\ -0.20}$ &  $1.10\substack{+0.06 \\ -0.05}$ &  $0.26\substack{+0.01 \\ -0.01}$ &  $0.010\substack{+0.002 \\ -0.001}$ &   $5.1\substack{+1.3 \\ -1.5}$ &  $6106\substack{+94 \\ -80}$ &  $1.40\substack{+0.04 \\ -0.04}$ &   $85.8\substack{+5.6 \\ -4.3}$ &          $-0.30\substack{+0.08 \\ -0.08}$ \\
  KIC10732098 &  $1.69\substack{+0.06 \\ -0.09}$ &  $1.14\substack{+0.03 \\ -0.04}$ &  $0.26\substack{+0.01 \\ -0.01}$ &  $0.017\substack{+0.002 \\ -0.002}$ &   $7.4\substack{+0.5 \\ -0.5}$ &  $5715\substack{+61 \\ -61}$ &  $1.77\substack{+0.04 \\ -0.03}$ &   $62.3\substack{+1.8 \\ -1.8}$ &           $0.07\substack{+0.06 \\ -0.07}$ \\
\bottomrule
\end{tabular}

%% file: tables/hyperparam_results.tex
\begin{tabular}{lccccc}
\toprule
Model &              $\Delta Y/\Delta Z$ &                             $\sigma_Y$ &                        $\mu_\alpha$ &                     $\sigma_\alpha$ &               $\alpha_\mathrm{mlt}$ \\
\midrule
            NP &  $1.69\substack{+0.21 \\ -0.21}$ &  $0.0074\substack{+0.0026 \\ -0.0022}$ &  $1.954\substack{+0.040 \\ -0.041}$ &  $0.065\substack{+0.030 \\ -0.024}$ &                                 --- \\
            PP &  $1.60\substack{+0.45 \\ -0.42}$ &  $0.0051\substack{+0.0045 \\ -0.0027}$ &  $1.742\substack{+0.081 \\ -0.070}$ &  $0.056\substack{+0.051 \\ -0.030}$ &                                 --- \\
           PPS &  $1.05\substack{+0.28 \\ -0.25}$ &  $0.0045\substack{+0.0038 \\ -0.0023}$ &  $1.900\substack{+0.095 \\ -0.088}$ &  $0.133\substack{+0.057 \\ -0.047}$ &                                 --- \\
            MP &  $1.60\substack{+0.45 \\ -0.42}$ &  $0.0051\substack{+0.0044 \\ -0.0027}$ &                                 --- &                                 --- &  $1.728\substack{+0.077 \\ -0.066}$ \\
           MPS &  $0.76\substack{+0.24 \\ -0.27}$ &  $0.0049\substack{+0.0039 \\ -0.0025}$ &                                 --- &                                 --- &  $2.088\substack{+0.031 \\ -0.029}$ \\
\bottomrule
\end{tabular}

%% file: tables/standardisation.tex
\begin{tabular}{lcccccccccc}
\toprule
{} & \multicolumn{5}{c}{Input} & \multicolumn{5}{c}{Output} \\
{} & $f_\mathrm{evol}$ & $M\,(\si{\solarmass})$ & $\mlt$ & $Y_\mathrm{init}$ & $Z_\mathrm{init}$ & $\log(\tau/\si{\giga\year})$ & $\teff\,(\si{\kelvin})$ & $R\,(\si{\solarradius})$ & $\dnu\,(\si{\micro\hertz})$ & $\metallicity_\mathrm{surf}\,(\si{\dex})$ \\
\midrule
$\mu_{1/2}$ &             0.865 &                  1.000 &  1.900 &             0.280 &             0.017 &                        0.790 &                5566.772 &                    1.224 &                     100.720 &                                     0.081 \\
$\sigma$    &             0.651 &                  0.118 &  0.338 &             0.028 &             0.011 &                        0.467 &                 601.172 &                    0.503 &                      42.582 &                                     0.361 \\
\bottomrule
\end{tabular}

%% file: tables/test_random.tex
\begin{tabular}{lcccccc}
\toprule
Error &  $\delta \tau/\tau\,(\%)$ &  $\delta T_\mathrm{eff}\,(\mathrm{K})$ &  $\delta R/R\,(\%)$ &  $\delta L/L\,(\%)$ &  $\delta \Delta\nu\,(\mathrm{\mu Hz})$ &  $\delta [\mathrm{M}/\mathrm{H}]_\mathrm{surf}\,(\mathrm{dex})$ \\
\midrule
$\mu_{1/2}$           &                    -0.012 &                                  0.070 &              -0.011 &               0.053 &                                0.00022 &                                                         0.00010 \\
$\sigma_\mathrm{MAD}$ &                     0.123 &                                  0.941 &               0.045 &               0.099 &                                0.06341 &                                                         0.00035 \\
\bottomrule
\end{tabular}

%% file: tables/sun_outputs.tex
\begin{tabular}{lccccccccc}
\toprule
                  $f_\mathrm{evol}$ &              $M\,(\si{\solarmass})$ &                           $\mlt$ &                   $Y_\mathrm{init}$ &                      $Z_\mathrm{init}$ &     $\tau\,(\si{\giga\year})$ &      $\teff\,(\si{\kelvin})$ &            $R\,(\si{\solarradius})$ &        $\dnu\,(\si{\micro\hertz})$ & $\metallicity_\mathrm{surf}\,(\si{\dex})$ \\
\midrule
 $0.517\substack{+0.009 \\ -0.008}$ &  $1.000\substack{+0.001 \\ -0.001}$ &  $2.12\substack{+0.03 \\ -0.03}$ &  $0.262\substack{+0.002 \\ -0.002}$ &  $0.0150\substack{+0.0003 \\ -0.0003}$ &  $4.6\substack{+0.1 \\ -0.1}$ &  $5777\substack{+12 \\ -12}$ &  $1.001\substack{+0.001 \\ -0.001}$ &  $135.37\substack{+0.14 \\ -0.14}$ &           $0.00\substack{+0.01 \\ -0.01}$ \\
\bottomrule
\end{tabular}